\documentclass{emulateapj}
\begin{document}

\title{SIM PlanetQuest Key 
Project Precursor Observations to Detect Gas Giant Planets Around Young Stars}

\author{Angelle Tanner\altaffilmark{1}, 
Charles Beichman\altaffilmark{1}, 
Rachel Akeson\altaffilmark{1},
Andrea Ghez\altaffilmark{2},
Konstantin N. Grankin\altaffilmark{3},
William Herbst\altaffilmark{4},
Lynne Hillenbrand\altaffilmark{5}, 
Marcos Huerta\altaffilmark{6},
Quinn Konopacky\altaffilmark{2},
Stanimir Metchev\altaffilmark{2}, 
Subhanjoy Mohanty\altaffilmark{7}, 
L. Prato\altaffilmark{8},
Michal Simon\altaffilmark{9}}

\altaffiltext{1}{Michelson Science Center, Caltech, 770 S. Wilson Ave., Pasadena, CA 91125 }
\altaffiltext{2}{UCLA, Division of Astronomy, 430 Portola Plaza, Box 951547, Los Angeles, CA 90095-1547 }
\altaffiltext{3}{Astronomical Institute of the Uzbek Academy of Sciences,
Tashkent, Uzbekistan, 700052}
\altaffiltext{4}{Astronomy Department, Wesleyan University, Middletown, CT 06459. }
\altaffiltext{5}{Caltech, MS 320-47, Pasadena, California 91125}
\altaffiltext{6}{University of Florida, 211 Bryant Space Science Center, Gainesville, FL, 32611-2055}
\altaffiltext{7}{Center for Astrophysics, Harvard University, 60 Garden Street Cambridge, MA 02138}
\altaffiltext{8}{Lowell Observatory, 1400 W. Mars Hill Rd., Flagstaff, AZ 86001}
\altaffiltext{9}{State University of New York at Stony Brook Dept. of Physics and Astronomy, Stony Brook, NY 11794-3800}

\begin{abstract}

We present a review of precursor observing programs for 
the SIM PlanetQuest Key project devoted to detecting Jupiter mass planets 
around young stars. In order to ensure that the stars in the sample
are free of various sources of astrometric noise that might impede the detection 
of planets, we have initiated programs to collect photometry, high contrast 
images, interferometric data and radial velocities for stars in both the Northern and Southern 
hemispheres. We have completed a high contrast imaging survey of target stars 
in Taurus and the Pleiades and found no definitive common proper motion 
companions within 
one arcsecond (140 AU) of the SIM targets. 
 Our radial velocity surveys have shown that many of
the target stars in Sco-Cen are fast rotators and a few stars in Taurus and
the Pleiades may have sub-stellar companions. Interferometric data
of a few stars in Taurus show no signs of stellar or 
sub-stellar companions with separations of $<$5 mas. The photometric survey suggests that approximately half 
of the stars initially selected for this program are variable to a degree 
($1\sigma>$0.1 mag) that would degrade the astrometric accuracy achievable 
for that star. While the precursor programs are still a work in 
progress,
we provide a comprehensive list of all targets ranked according to their 
viability
as a result of the observations taken to date. By far, the observable that
removes the most targets from the SIM-YSO program is photometric variability.

\end{abstract}

\keywords{astrometry --- stars: pre-main sequence --- extrasolar planets}

\section{Introduction}

The majority of the over 200 planets found to date have been detected using either 
radial velocity (RV) or transit studies in orbits ranging from less than 0.1 AU out to  
beyond 5 AU, with a wide range of eccentricities, and masses ranging from less than 
that  of Uranus up to many times that of Jupiter (Butler et al. 2006). However, the 
host stars of these planets are mature main sequence stars which were chosen based 
on their  having quiescent photospheres for the successful measurement of small 
Doppler velocities  ($<$10 m s$^{-1}$). Similarly, stellar photospheres must be 
quiescent at the  milli-magnitude level for transit detections since a Jupiter 
mass planet transiting a solar type 
star reduces the photometric signal by about 1.4\%. Since young stars often have radial 
velocity fluctuations or rotationally broadened line widths of {\it at least} 
500 m s$^{-1}$ and brightness fluctuations of many percent, RV measurements 
accurate to $<$100 m s$^{-1}$ or transit observations cannot be used to detect 
planets around young stars\footnote{A number of groups are attempting RV 
observations in the near-IR since at these wavelengths it may be possible to 
improve on these limits and find a few ``hot Jupiters'' within 0.1 AU}. A few 
potentially planetary mass objects have been detected at 20-100 AU from young
 host stars ($<$10 Myr) by direct, coronagraphic 
imaging, e.g 2MASSW J1207334-393254 (Chauvin et al. 2005) and GQ Lup 
(Neuh{\"a}user et al. 2005). However, these companions are only inferred to be of 
planetary mass by comparison to uncertain evolutionary models that predict the 
brightness of ``young Jupiters'' as a function of mass and age 
(Wuchterl \& Tscharnuter 2003; Baraffe et al. 2003; Burrows et al. 1997). 
Since dynamical determinations of mass are impossible for objects on such 
distant orbits, it is difficult to be sure that these are planets and not 
brown dwarfs. Nor is it even clear than the origin of these distant 
``young Jupiters'' is due to same formation processes as planets found 
closer-in. Multiple fragmentation events (Boss 2001), rather than core accretion 
in a dense disk (Ida \& Lin 2004), may be responsible for the formation of 
these distant objects. As a result of the selection biases of the radial velocity, 
transit and direct imaging 
techniques, we know little about the incidence of close-in planets around young 
stars, leaving us with many questions about the formation and 
evolution of gas giant planets. 

Given the observational limitations and uncertainties that are inherent 
to radial velocity and direct imaging, micro-arcsecond astrometry 
is a feasible and direct method for estimating the masses of giant planets
around stars in young clusters which lie at distances closer than 140 parsecs
(Beichman et al. 2001). Equation (1) gives the astrometric amplitude, $\phi$, in 
units appropriate to the search for planets around young stars:

$$ \phi = 35 \, { {140 pc} \over {D_{pc}}} \, { {a_{AU}} 
\over {5.2 AU}} \, { {M_p} \over {M_{J}}} \, { { M_\odot} \over {M_\star} } \, \, 
\mu{\rm as} \,\,\,\, (1) $$

\noindent where 
$\phi$ is the astrometric amplitude in $\mu$as, 
$D$ is the distance to the star in parsecs, 
$a$ is the orbital semi-major axis of the planet in AU, 
$M_p$, is the planet mass in Jupiter masses, 
and $M_\star$ is the star's mass in solar masses. 
Thus, a Jupiter orbiting 5.2 AU away from a 0.8 $M_\odot$ star at the distance of 
youngest stellar associations (1-10 Myr) such as Taurus and Chamaeleon 140 pc away 
would produce an astrometric amplitude of 44 $\mu$as. At the 25-50 pc distance of 
the nearest young stars (10-50 Myr) such as members of the $\beta$ Pic and TW Hya 
moving groups, the same system would have an astrometric amplitude in excess of 100 $\mu$as.
Moving a Jupiter into a 1 AU orbit would reduce the 
signal by a factor of 5.2, or 50 $\mu$as for a star at 25 pc and 8 $\mu$as for one in 
Taurus. Table~\ref{clusttab} lists the star formation regions and young moving groups being 
included in the SIM-YSO survey, a SIM PlanetQuest key project aimed at
detecting Jupiter-mass planets around young stars. Since SIM will be able to detect astrometric signals with a Single Measurement 
Accuracy (SMA) of 4 $\mu$as (1$\sigma$) in a fairly quick ``Narrow Angle (NA)'' 
observation and 11 $\mu$as in single ``Wide Angle (WA)'' observation, a search 
for gas giants falls well within SIM's capabilities for wide and narrow angle 
astrometry and forms the core of the SIM-YSO program. With SIM's sensitivity 
it is reasonable to study stars brighter than R$\sim$12 mag to a level such 
that the expected astrometric amplitude of 8 $\mu$as for stars at 140 pc is 
detected with  2$\sigma$ confidence in each measurement. This astrometric 
accuracy is appropriate for detecting planets of unknown orbital parameters 
with a series of approximately 75-100 1-D measurements (Sozzetti et al. 2003; 
Catanzarite et al. 2006). 

\begin{figure}[ht]
\plotone{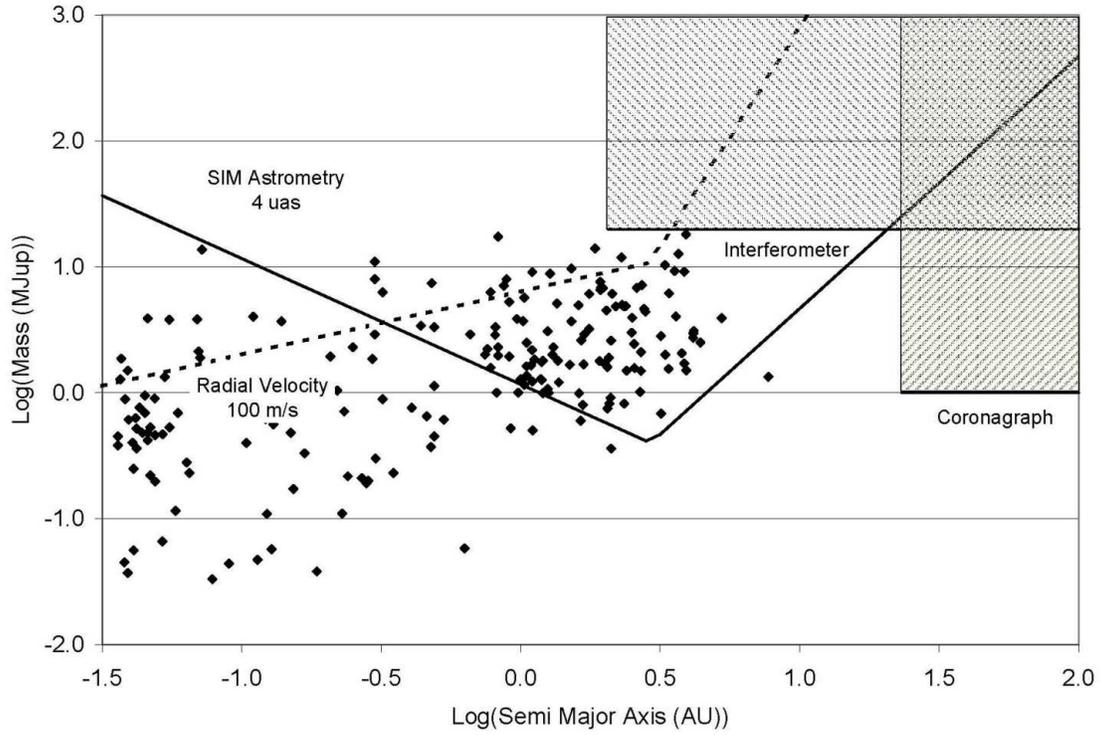} 
\figcaption{ Plot of planet mass in M$_J$ versus semi-major
axis of the sensitivities expected from the SIM-YSO survey (solid-line) in
addition to ground based coronagraph (labeled), interferometry (labeled) and radial velocity (dashed
line) surveys
of young stars. Also plotted are the properties of the known 
radial velocity planets (diamonds). All these sensitivity limits assume a distance of 140 pc. \label{simperform}}
\end{figure}

Figure~\ref{simperform} shows orbital location (semi-major axis) and  Msin(i) for over
200 known planets orbiting nearby mature stars\footnote{Planet information from Schneider (2007)}.
These planets were found using radial velocity measurements with noise levels
as low as 1 m s$^{-1}$. However, for the the young stars considered
here, the radial velocity measurements will be limited, even in the  near-infrared,
to 100-500 m s$^{-1}$ (or greater) due to rapid rotation, veiling, and photospheric variability.
Thus, we plot a RV sensitivity curve for planets orbiting a 1 M$_\sun$ star assuming
a limiting accuracy of 100 m s$^{-1}$. For comparison, we plot the astrometric sensitivity curve
for our SIM-YSO project where we demand that  the minimum detectable planet have
an amplitude twice the single measurement accuracy of 4 $\mu$as. The curve is plotted for a
1 M$_\sun$ star located at the distance of Taurus (140 pc).  For planets with periods greater than
the nominal mission duration of 5 years, we have degraded the sensitivity as shown schematically
in the plot. Finally, we estimate how a coronagraph on a 30-m telescope or an interferometer  with an 85 m baseline, both
operating at 1.6 $\mu$m and  reaching down to the Jovian mass range at orbital distances of 10 AU
or greater would complement, but not replace, SIM observations. 

In order to maximize the scientific yield of this SIM key project, a significant
effort is required to gather information about the target stars prior to the 
launch of SIM. A careful vetting of the target list is required to reject stars that might be problematic due to the presence of starspots that
 might induce large astrometric offsets, to the presence of circumstellar 
emission from scattered light, or the presence of either 
visible and spectroscopic companions. 

To ensure that we observe astrometrically stable systems, we have initiated 
a series of precursor observations to check for nearby infrared companions ($\S$3), 
radial velocity variations due to unseen companions ($\S$4), and photometric variability ($\S$5). 
Table~\ref{progtab} summarizes those precursor programs presently being conducted 
along with the telescope and the principal investigator. In subsequent sections we 
summarize the different programs and give detailed results from a number of them. 
The SIM-YSO target list will continue to evolve as we add new stars for their scientific 
interest and remove stars due to one failing or another. In the end, we intend to
have a complete sample in both stellar age and mass that will allow for a 
statistically significant study of planets around young stars. 
 
This paper describes the present status of the SIM-YSO sample, the overall strategy 
for the precursor vetting program, and detailed results of one specific program, 
the Palomar AO survey aimed at identifying companions to the SIM-YSO targets in 
the Pleiades and Taurus. We will summarize briefly  the progress being made on our 
photometric and radial velocity surveys of both the Northern and Southern targets 
and their implications for the SIM-YSO target sample.

\section{The SIM-YSO Stellar Sample of Young Stars}

In a survey of $\sim$200 young stars we expect to find anywhere from 10-20 
to 200 planetary systems depending on whether the 5-10\% of stars with
known radial velocity planets are representative of the younger planet population or 
whether all young stars have planets only to lose them to inward migration.  The youngest 
stars in the sample (see Table~\ref{clusttab}) will be located in well known star-forming 
regions and will be observed in Narrow Angle mode which is capable of 
achieving single measurement accuracies of 4 $\mu$as. Somewhat older stars, 
such as those in the $\beta$ Pictoris 
and TW Hydrae Associations, are only 25-50 pc away, and can be observed less 
expensively in Wide Angle mode capable of a single measurement accuracy of 
$<$11 $\mu$as (Unwin 2005).

We have set our sensitivity threshold to ensure the detection of Jupiter-mass 
planets in the critical orbital range of 1 to 5 AU. These observations, when 
combined with the results of the SIM planetary searches of mature 
stars, will allow us to test theories of planetary formation and early solar 
system evolution. By searching for planets around pre-main sequence stars 
carefully selected to span an age range from 1 to 100 Myr, we will learn at what 
epoch and with what frequency giant planets are found at the water-ice ``snowline" 
where they are expected to form (Pollack et al 1996). This will provide insight 
into the physical mechanisms by which planets form and migrate from their place 
of birth, as well as their survival rate. With these observations in hand, we 
will provide data, for the first time, on such important questions as: What 
processes affect the formation and dynamical evolution of planets? When and 
where do planets form? 
What is initial mass distribution of planetary systems around young stars? How 
might planets be destroyed? What is the origin of the eccentricity of planetary 
orbits? What is the origin of the apparent dearth of companion objects between 
planets and brown dwarfs seen in mature stars? How might the formation and 
migration of gas giant planets affect the formation of terrestrial planets?

Our observational strategy is a compromise between the desire to extend the 
planetary mass function as low as possible and the essential need to build up 
sufficient statistics on planetary occurrence. About half of the sample will be 
used to address the ``where" and ``when" of planet formation. We will study 
classical T Tauri stars (cTTs) which have massive accretion disks as well as 
post-accretion, weak-lined T Tauri stars (wTTs). Preliminary estimates suggest 
the sample will consist of $\sim$30\% cTTs and $\sim$70\% wTTs, driven in part 
by the difficulty of making accurate astrometric measurements toward objects 
with strong variability or prominent disks. The extent to which this distribution of 
cTTs and wTTs survives the screening programs for photometric and
dynamic stability will be addressed in $\S$6.  The second half of the sample will 
be drawn from the closest, young clusters with ages starting around 5 Myr
to the 10 Myr thought to mark the end of prominent disks, and ending around the 100 
Myr age at which theory suggests that the properties of young planetary systems 
should become indistinguishable from those of mature stars. The properties of 
the planetary systems found around stars in these later age bins will be used 
to address the effects of dynamical evolution and planet destruction (Lin et al. 2001).

 We have adopted the following criteria in developing our 
initial list of candidates: a) stellar mass between 0.2 and 2.0 M$_\odot$; 
b) R $< 12$ mag for reasonable integration times; 
c) distance less than 140 pc to ensure an astrometric signal greater 
than 6 $\mu$as; d) no companions within 2$^{\prime\prime}$ or 100 AU for 
instrumental and scientific considerations, respectively; 
e) no nebulosity to confuse the astrometric measurements; 
f) variability $\Delta R<$0.1 mag; and
g) a spread of ages between 1 Myr and 100 Myr to encompass the 
expected time period of planet-disk and early planet-planet interactions. With 
proper selection, the effect of various astrophysical disturbances can be kept 
to less than the few $\mu$as needed to detect Jupiter-mass planets at $\sim$50-140 pc.

\begin{figure}[ht]
\plotone{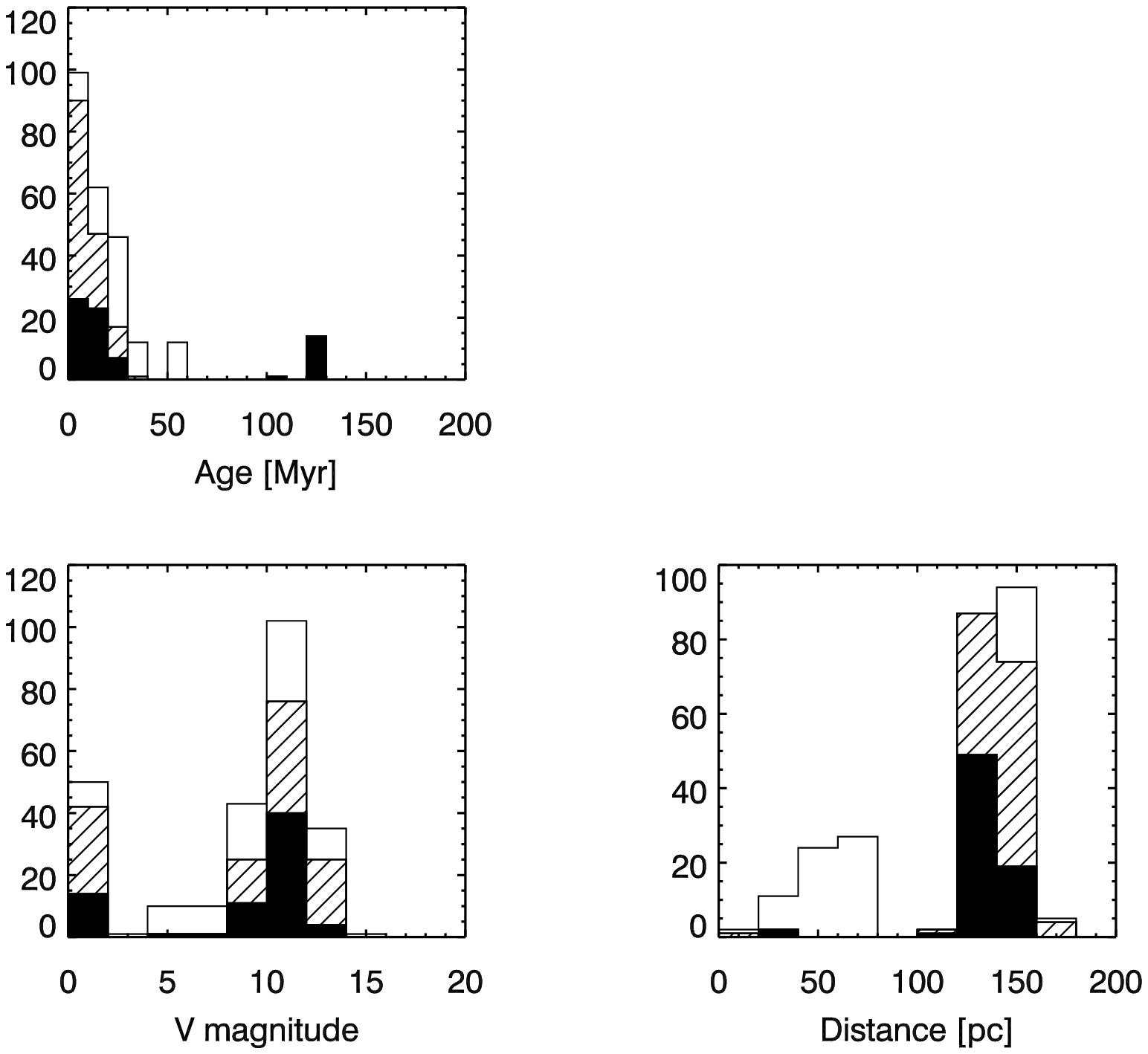} 
\figcaption{Histograms of the properties of the SIM-YSO sample. The slashed and
black bars show those targets potentially eliminated and 
remaining, respectively, after cuts from photometry or nearby companions.
The white bars represent those targets yet to be observed in the precursor
surveys. \label{samhist}}
\end{figure}

The initial SIM-YSO sample (see Table~\ref{samtab}) consists of stars in the well known star-forming 
regions and close associations. Figure~\ref{samhist} 
shows histograms of the properties of the stars in the sample including distance, 
V magnitude and age.  The stars included in the initial sample have been screened 
for binarity in either  imaging (Stauffer et al. 1998; Lowrance et al. 2005) or 
spectroscopic surveys (White \& Ghez 2001; Mathieu et al. 1997; Steffen et al. 2001). 

\begin{center}
\begin{deluxetable}{lccc}
\tablecaption{SIM-YSO Sample \label{clusttab}}
\tablehead{ \colhead{Cluster}  & \colhead{Age} & \colhead{Distance} & \colhead{\# stars} \\
                               & \colhead{[Myr]} & \colhead{[pc]}       &
}
\startdata
 Beta Pic  & 20   & 10-50 & 16\\
 Chameleon & 1-10 & 140   & 8 \\
 Eta Cha   & 4-7  & 100   & 2 \\
 Horologium & 30  &  60   & 12\\
 IC2391     & 53  & 155   & 12 \\
 Ophiuchus  & 2   & 160   &  5\\
 Pleiades   & 125 & 130   & 14\\
 TW Hydra   & 10  &  60   & 15\\
 Taurus-Aureiga & 2 & 140 & 25\\
 Tucanae    & 20 & 45     & 20\\
 Upper Sco  & 1-10 & 145  & 49\\
 Sco Cen    & 1-25 & 130  & 81\\
\enddata
\end{deluxetable}
\end{center}

\begin{center}
\begin{deluxetable}{lcc}
\tablewidth{0pt}
\tablecaption{Precursor Programs \label{progtab}}
\tablehead{ \colhead{Program}  & \colhead{Telescope}& \colhead{PI} 
}
\startdata
AO Imaging (North) & Palomar & Tanner\\
AO Imaging (South) & VLT     & Dumas \\
Speckle Imaging (North) & Keck & Ghez/Konopacky \\
V$^2$ (North) & Keck & Akeson \\
RV Survey  (North) & McDonald & Prato \\
RV Survey  (South) & CTIO/Magellan & Mohanty \\
Photometry (North) & Maidanak  & Grankin \\
Photometry (South) & SMARTS     & Simon \\
\enddata
\end{deluxetable}
\end{center}

\medskip

\section{High Contrast Direct Imaging}

We begin by presenting the results of a companion survey to those SIM-YSO 
targets in our Taurus and Pleiades samples (see Table~\ref{sample}). Companions 
within the
1$\farcs$5 field-of-view of the SIM interferometer which have magnitudes within $\Delta$V$\sim$ 
4 (M. Shao, private comm.) could cause a bias in the position 
of the fringe used to make 
the astrometric measurements. Additionally, a massive, unknown stellar companion will 
induce astrometric perturbations complicating the 
astrometric solution for a planet around the primary star.  To look for 
common proper motion companions to the SIM-YSO stars,  we have conducted an 
adaptive optics (AO) coronagraphic imaging survey around 31 stars in the 
Taurus (2 Myr, 140 pc, Kenyon et al. 1994) and Pleiades  
(120 Myr, 135 pc, Stauffer et al. 1998; Pan et al. 2004)  
clusters with the PALAO adaptive optics system and its 
near-infrared camera, PHARO, on the 
Hale 200-inch telescope at Palomar (Hayward et al. 2001). 
These data will reveal the presence of stellar and brown dwarf companions 
located between $\sim$50 and 1000 AU, and, in the case of the youngest 
stars in these systems ($\sim$2 Myr), will be sensitive to hot, young 
planets with masses in  the range of 10-20 M$_J$ (Burrows et al. 1997; 
Baraffe et al. 2003). 

In the process of searching for unseen companions around these stars, we 
are also addressing planet formation issues. By investigating whether the 
``brown dwarf desert'' observed for separations of $<$5 AU around main 
sequence stars (Marcy et al. 2000) also exists at larger separations for 
young stars, we can test whether brown dwarfs are formed at this separation 
and subsequently migrate inward and are destroyed by falling onto the star. 
In this case we might find that T-Tauri stars have a larger population of 
brown dwarfs than main sequence 
stars at these separations. 

\begin{center}
\begin{footnotesize}
\begin{deluxetable}{lcccccc}
\tablecaption{Palomar Target Sample\label{sample}}
\tablehead{ \colhead{Target}  & \colhead{V}      & \colhead{2MASS K$_s$} &\colhead{SpTy} & \colhead{Age}   & \colhead{Distance} \\
            \colhead{}        & \colhead{[mag]}  & \colhead{[mag]}       &\colhead{}     & \colhead{[Myr]} & \colhead{ [pc]}
}
\startdata
HII 1032  &  11.1   &  9.16$\pm$0.02  &  G8  &  125  &  130  \\
HII 1095  &  11.92  &  9.67$\pm$0.02  &  K0  &  125  &  130  \\
HII 1124  &  12.12  &  9.86$\pm$0.02  &  K1  &  125  &  130  \\
HII 1136  &  12.02  & 12.14$\pm$0.02  &  G8  &  125  &  130  \\
HII 1275  &  11.47  &  9.53$\pm$0.02  &  G8  &  125  &  130  \\
HII 1309  &  9.58   &  8.28$\pm$0.02  &  F6  &  125  &  130  \\
HII 1514  &  10.48  &  8.95$\pm$0.02  &  G5  &  125  &  130  \\
HII 1613  &  9.87   &  8.57$\pm$0.02  &  F8  &  125  &  130  \\
HII 1794  &  10.2   &  8.89$\pm$0.02  &  F8  &  125  &  130  \\
HII 1797  &  10.09  & 15.04$\pm$0.11  &  F9  &  125  &  130  \\
HII 1856  &  10.2   &  8.66$\pm$0.02  &  F8  &  125  &  130  \\
HII 2366  &  11.53  &  9.55$\pm$0.02  &  G2  &  125  &  130  \\
HII 430   &  11.4   &  9.47$\pm$0.02  &  G8  &  125  &  130  \\
HII 489   &  10.38  &  8.87$\pm$0.02  &  F8  &  125  &  130  \\
AA Tau    &  12.82  &  8.05$\pm$0.02  &  K7  &  2  &  140  \\
BP Tau    &  11.96  &  7.74$\pm$0.02  &  K7  &  2  &  140  \\
DL Tau    &  13.55  &  7.96$\pm$0.02  &  G  &  2  &  140  \\
DM Tau    &  13.78  &  9.52$\pm$0.02  &  K5  &  2  &  140  \\
DN Tau    &  12.53  &  8.02$\pm$0.02  &  M0  &  2  &  140  \\
DQ Tau    &  13.66  &  7.98$\pm$0.02  &  M0  &  2  &  140  \\
DR Tau    &  13.6   &  6.87$\pm$0.02  &  K5  &  2  &  140  \\
GK Tau    &  12.5   &  7.47$\pm$0.02  &  K7  &  2  &  140  \\
IP Tau    &  13.04  &  8.35$\pm$0.02  &  M0  &  2  &  140  \\
IQ Tau    &  14.5   &  7.78$\pm$0.02  &  M0.5  &  2  &  140  \\
IW Tau    &  12.51  &  8.28$\pm$0.03  &  K7  &  2  &  140  \\
LkCa 19   &  10.85  &  8.15$\pm$0.02  &  K0  &  2  &  140  \\
V1072 Tau &  10.3   &  8.30$\pm$0.02  &  K1  &  2  &  140  \\
V830  Tau &  12.21  &  8.42$\pm$0.02  &  K7  &  2  &  140  \\
V836 Tau  &  13.13  &  8.60$\pm$0.02  &  K7  &  2  &  140  \\
\enddata 
\end{deluxetable}
\end{footnotesize}
\end{center}

\subsection{Data Reduction and Analysis}

The Palomar observations were obtained over three observing runs - 
Oct. 23, 2003; Dec. 4-6 2003; and Nov. 12-14 2005 - with
good (0$\farcs$2) to moderate (0$\farcs$5) seeing throughout the nights. 
The PHARO/PALAO camera has a pixel 
scale of 25 mas pixel$^{-1}$ and a field-of-view of 25 arcseconds. Each target was observed with 
the 0$\farcs$97 diameter occulting spot placed over the star with integration times 
of 60 seconds each and multiple (10-20) images collected per target. Sky images 
were also taken adjacent to each set of target images by offsetting 
30$^{\prime\prime}$ from the target in the four cardinal directions. 
For flux calibration, observations of the target stars were taken with 
the star offset from the coronagraph in a five point dither pattern to  
allow for adequate sky subtraction. To improve observing efficiency, 
those stars with similar magnitudes and colors were paired together to 
act as each other's point spread function. Two sets of known binary stars with high quality 
orbital solutions (WDC 09008+4148 and WDC 23052-0742) were also observed 
to provide an accurate determination of the plate scale and image orientation. 
During these observations, each binary was placed in multiple positions over the 
field of view of the camera.

\begin{figure}[ht]
\plotone{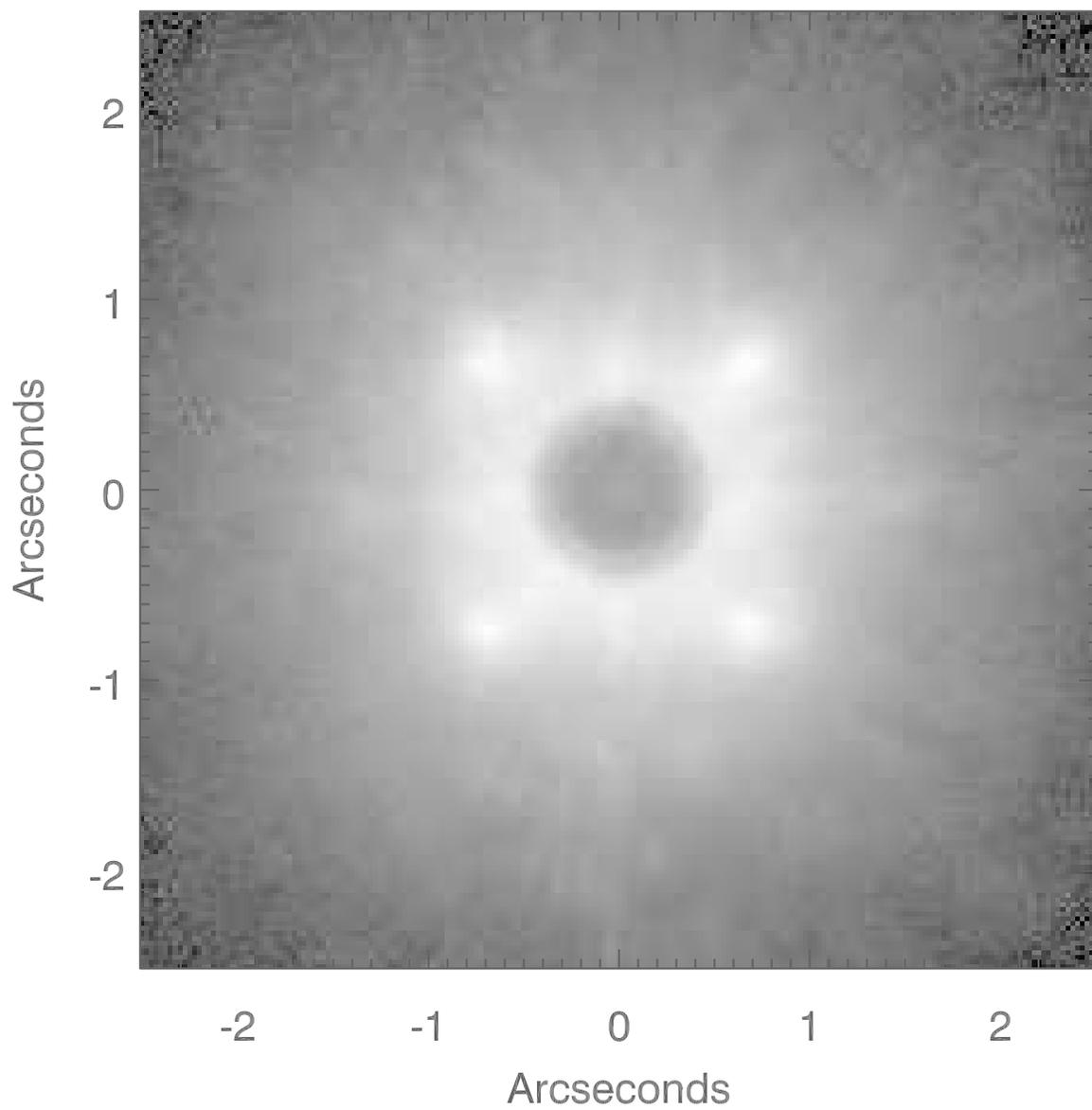}
\figcaption{ Stretched greyscale image of LkCA19 highlighting the poisson spot and
waffle pattern inherent in the Palomar AO+coronagraph PSF. North is up and East is
to the left. \label{lkcaim}}
\end{figure}

\begin{figure}[ht]
\plotone{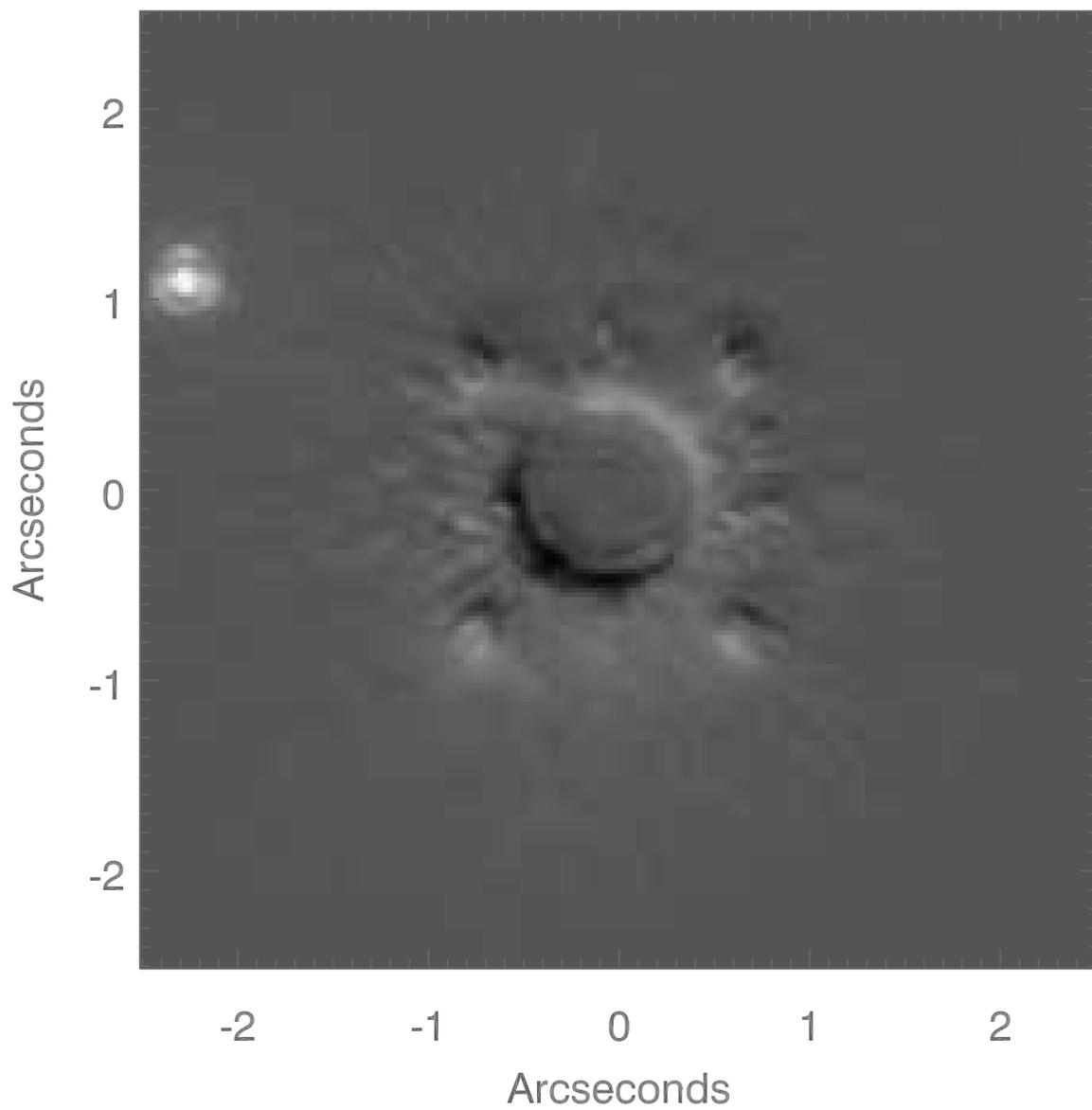}
\figcaption{Greyscale difference image of GK Tau and V830 Tau. The apparent companion to
GK Tau is 2.4" away but is most likely a background star due to its blue
colors and non-common proper motion. North is up and East is to the left.  \label{gkim}}
\end{figure}

All the images were sky-subtracted, flat-fielded and corrected for bad-pixels. 
We utilized a number of methods for image registration prior to PSF subtraction. 
The best method involved using the centroid of either the waffle pattern or the 
``Poisson" spot (Metchev 2005). Using these methods we were able to achieve a registration 
accuracy of 2.3 and 0.7 pixels, 
respectively, which is comparable to the accuracy achieved in a similar 
survey using the same instrument (Metchev et al. 2004). After registration, the 
median of each image stack was calculated to produce the final image 
(see Figure~\ref{lkcaim}). The images of those target stars with similar
magnitudes and colors are paired and subtracted to produce a difference image
intended to reduce the residual flux within one arcsecond of the coronagraphic spot. 
Prior to subtraction, each pair of images was scaled to have the same peak flux within 
the coronagraph halo to ensure a minimal residual flux after subtraction. 
Figure~\ref{gkim} shows the difference image of GK Tau and V830 Tau with a bright 
companion candidate next to GK Tau.  

A thorough visual inspection of both the median-averaged, coronagraphic images
and the difference images was performed to identify all potential 
companions. All visual companions identified in these images 
are listed in Table~\ref{comptab} along with their distance from the center of 
the coronagraph, position angle and magnitude difference compared to the target
star. The coronagraphic images have been flux calibrated using the images 
taken with the primary star off-set from the coronagraph while accounting
for a difference in integration time and the well-defined neutral density
filter used for the off-spot images (Metchev et al. 2004). 
The magnitudes for both the primary stars in the off-spot images
and the companions in the coronagraphic images are 
estimated from aperture photometry with an aperture of 0$\farcs$9 and sky 
annulus of 1$\farcs$1-1$\farcs$4. The K$_s$ band magnitudes of the primaries were 
taken from the Two Micron All-Sky Survey (Cutri et al. 2006). Since many of
the target stars are variable in the optical, the near-infrared photometric calibration may
be uncertain although these stars are typically 2-3 times less variable in the 
infrared than the visible (Eiroa et al. 2002).  

\subsection{Results}

\begin{figure}[ht]
\plotone{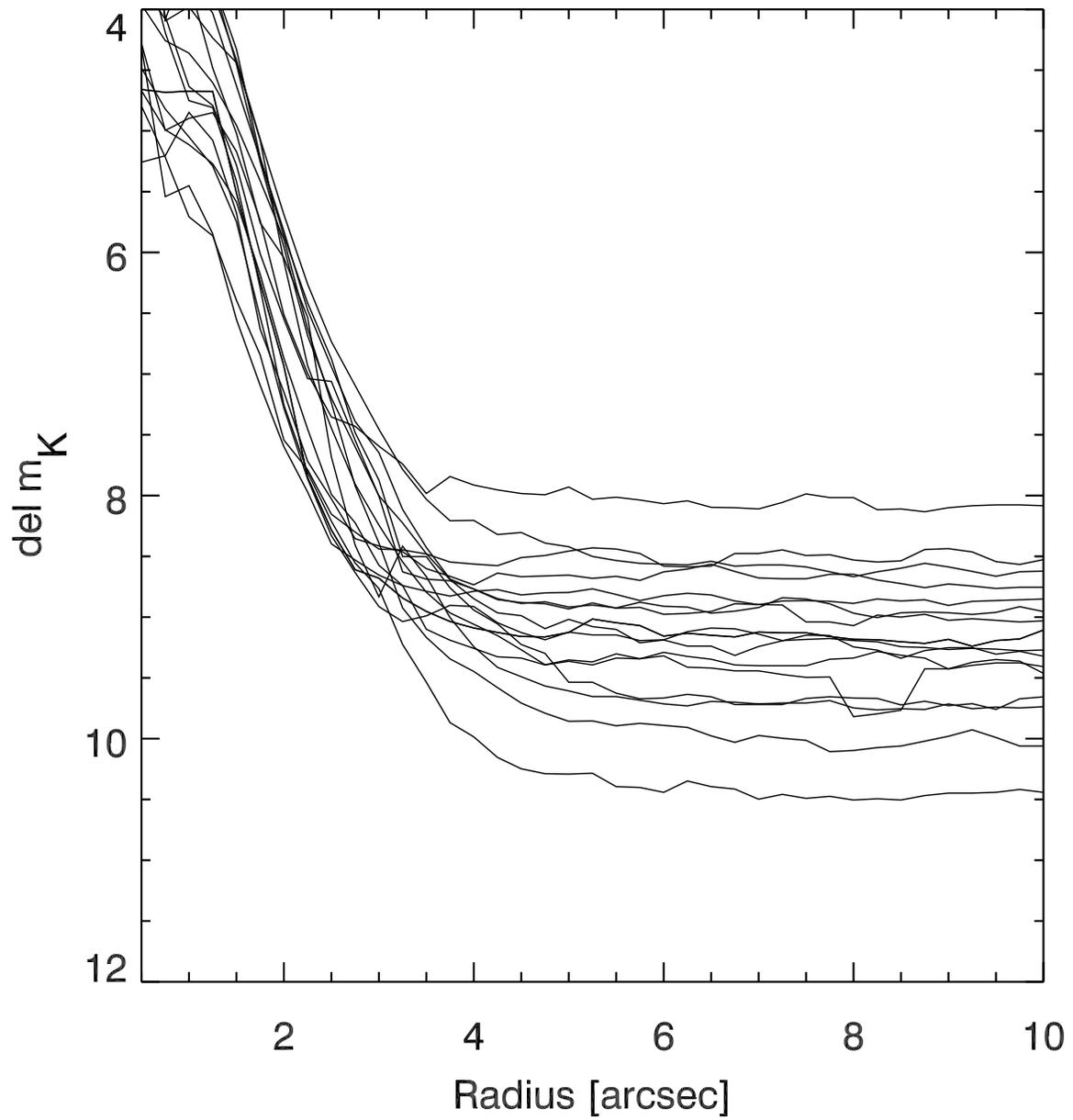}
\figcaption{Plot of the contrast in $\Delta K_s$  magnitude detectable in the PHARO images as a function of
separation from the target. \label{senplot}}
\end{figure}

In Taurus and the Pleiades, 10 out of 16 of the targets (63\%) and 5 out of 14 (36\%),
respectively, have visual companions within the 25$''$ ($\sim$1750 AU) field-of-view. 
Almost all of the companions lie $>$2 arcseconds away from the target 
corresponding to a distance greater than 300 AU. To estimate the sensitivities 
of our images as a function of distance from the star, we employ ``PSF planting"
in which a PSF corresponding to an object of known brightness is inserted into the 
image to determine whether it is detectable.  
A PSF extracted from the off-spot image of each target is sky subtracted, 
normalized, multiplied by an array of contrast values ($\Delta K_s$=7.7-15.1 mags) 
and placed at a 
range (0-5$''$) of distances from the target at random position
angles. We completed 10,000 iterations of the PSF planting algorithm 
to fill out the parameter space of contrast and distance from the primary star.  
To determine whether the planted star is detected, the image is cross-correlated 
with a flux normalized PSF. The correlation values are binned according to the distance of the 
PSF from the star in increments  of 0$\farcs$1. For each distance bin we 
estimate the minimum PSF intensities which resulted in a correlation value of 0.9 
or higher.  The intensities are converted into magnitudes using the flux 
calibration from the off-spot image and the 2MASS K$_s$ magnitude of the star. 
Figure~\ref{senplot} plots the largest K$_s$ magnitude difference 
between the target star and planted PSF with a
correlation of 0.9 as a function 
of distance from the star for all targets with calibration
data. Table~\ref{sentab} lists the values of the faintest detectable K$_s$ magnitudes 
at 0.5, 1, 2, 5 and 9 arcseconds. On average, we were
able to detect sources with a magnitude contrast 
of $\Delta K_S$ = 4-7 mag at 2$''$ and $\Delta K_S$$\sim$8-10 at 5$''$. The range 
of image contrasts is due primarily to variations in seeing conditions 
throughout the night since all the targets had similar magnitudes and integration 
times. 

\begin{deluxetable}{lccc}
\tablecaption{Visual Companions to Pleiades and Taurus Targets\label{comptab}}
\tablehead{ \colhead{Target}  & \colhead{Separation} & \colhead{PA}        &  \colhead{ K$_s$} \\
            \colhead{}        & \colhead{ ["]}       & \colhead{[degrees]} &  \colhead{[mag]}
}
\startdata
HII 1032  &  12.7  &   45.0  &  16.0    \\
HII 1309  &  11.2  &  -27.2  &  15.0    \\
HII 1309  &  12.1  &  108.8  &  15.2    \\
HII 1797  &   6.6  &   -8.8  &  15.7    \\
HII 489   &  13.4  &  144.4  &  17.1    \\
AA Tau    &   5.9  &   98.6  &  15.2    \\
BP Tau    &   3.1  &  -83.6  &  14.0    \\
BP Tau    &   5.6  &    8.0  &  15.0    \\
DL Tau    &  12.8  &   22.7  &  13.4    \\
DL Tau    &   8.5  &   61.8  &  14.5    \\
DL Tau    &  16.7  &  134.8  &  14.8    \\
DL Tau    &  11.9  &   15.0  &  15.4    \\
DQ Tau    &   7.3  &  149.6  &  15.3    \\
GI Tau    &   8.3  &    1.7  &  14.6    \\
GK Tau    &   2.4  &   62.1  &  12.1    \\
IP Tau    &   3.7  &  124.3  &  15.1    \\
IQ Tau    &   9.8  &  165.5  &  14.9    \\
IQ Tau    &  13.9  &  -64.6  &  15.3    \\
IQ Tau    &  10.7  &  -35.9  &  13.4    \\
IQ Tau    &  10.2  &  -33.9  &  13.6    \\
LkCa 19   &   4.3  &  -77.4  &  16.2    \\
LkCa 19   &  11.8  &  -36.2  &  13.9    \\
V830 Tau  &   7.1  &  -52.3  &  17.0    \\
V830 Tau  &  11.2  &   58.3  &  15.8    \\
V830 Tau  &  11.8  &  100.3  &  17.5    \\
V830 Tau  &   7.8  &  146.1  &  17.7    \\
\enddata
\end{deluxetable}

Calibration binaries were used to estimate the plate scale of the PHARO camera.  
For the Oct 2004 data, we adopt a plate scale of 25.11$\pm$0.04 mas pixel$^{-1}$ 
estimated from 3 different binary stars (WDS 09006+4147, WDS 18055+0230, 
WDS 20467+1607) that were observed very close to our observations using the 
same instrument (Oct 4-5, 2004, Metchev 2005). For the 2005 data we estimate 
a plate scale of 25.21$\pm$0.36 mas pixel$^{-1}$ .  
This plate scale and its uncertainty comes from the average and standard 
deviation of the separation of one binary (WDS 09006+4147) placed in multiple
positions across the field of view after correction for the known distortion in
the camera.   

To estimate the position of the occulted target star in these images, 
we use the waffle pattern inherent to every PSF (see Figure~\ref{lkcaim}). 
Each waffle pattern consists
of four points in a box pattern around the star. The center of the coronagraphic 
PSF is determined from the intersection of the two diagonal lines fitted to the centroid positions
of the four peaks in the waffle pattern. Using this method, we are able to determine the
position of the star to within 0.35 pixels estimated from the standard deviation
of the stellar position in a stack of sub-frames. 
The positions of the companion candidates are estimated
from their centroids and have errors of 0.5-1 pixels depending on the brightness 
of the companion, seeing conditions and telescope drift. The pixel positions for 
all primaries and their companions are corrected for the distortion determined for the 
PHARO camera (Metchev 2005). The errors in the stellar position and PHARO pixel scale are
propagated into the error of the offsets of the companion candidates from their 
primaries. 

\begin{figure}[ht]
\plottwo{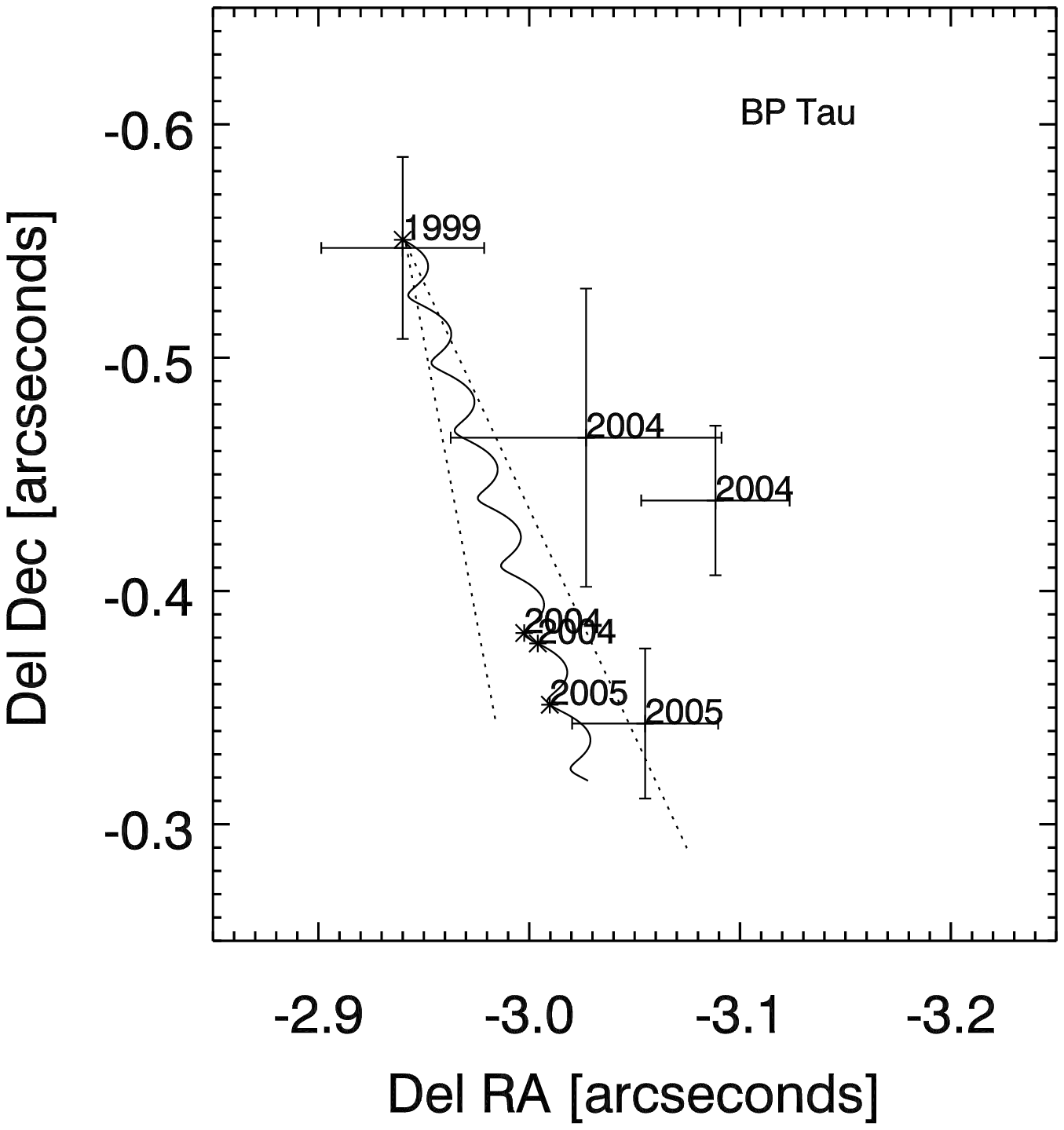}{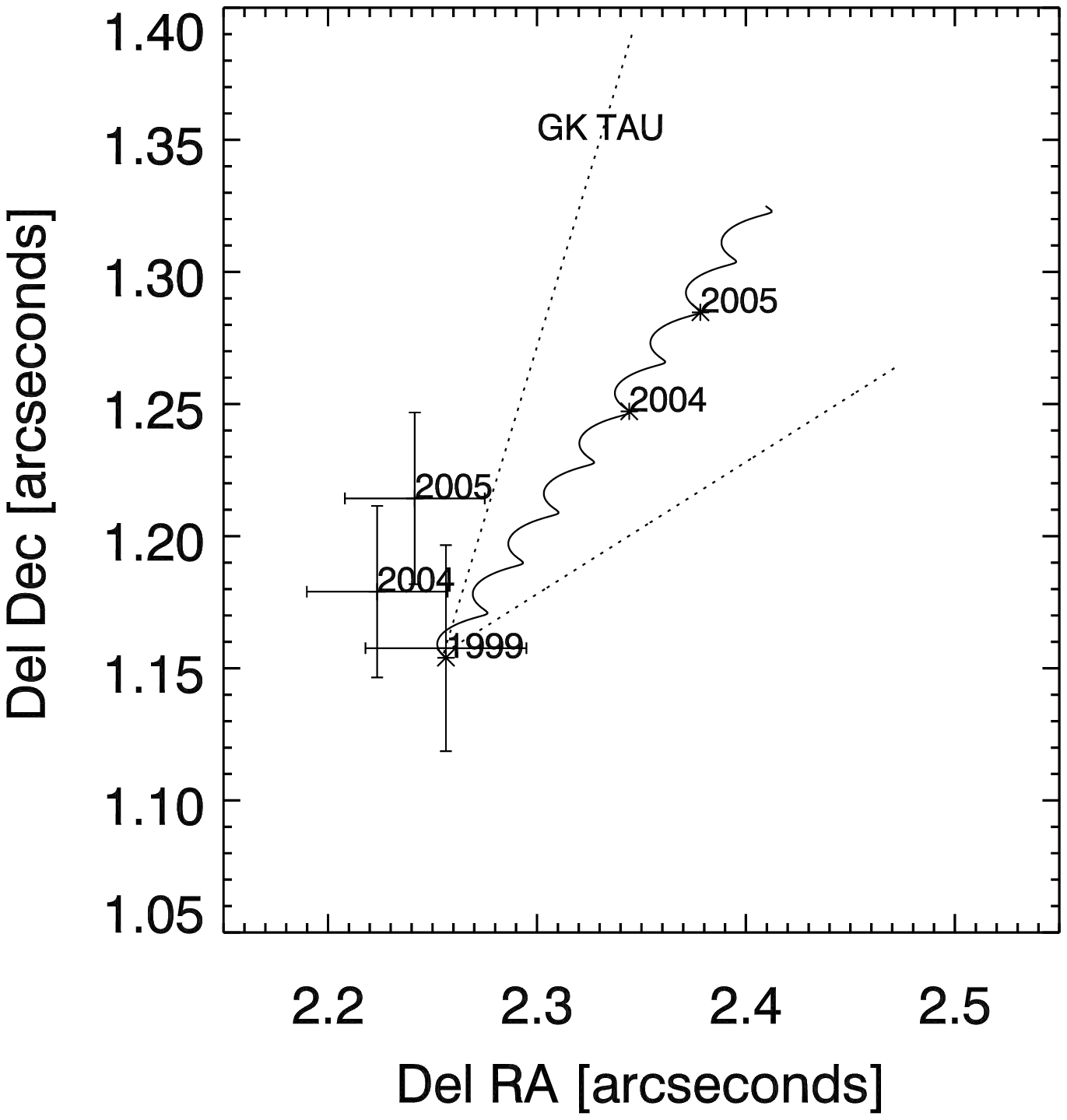}
\plottwo{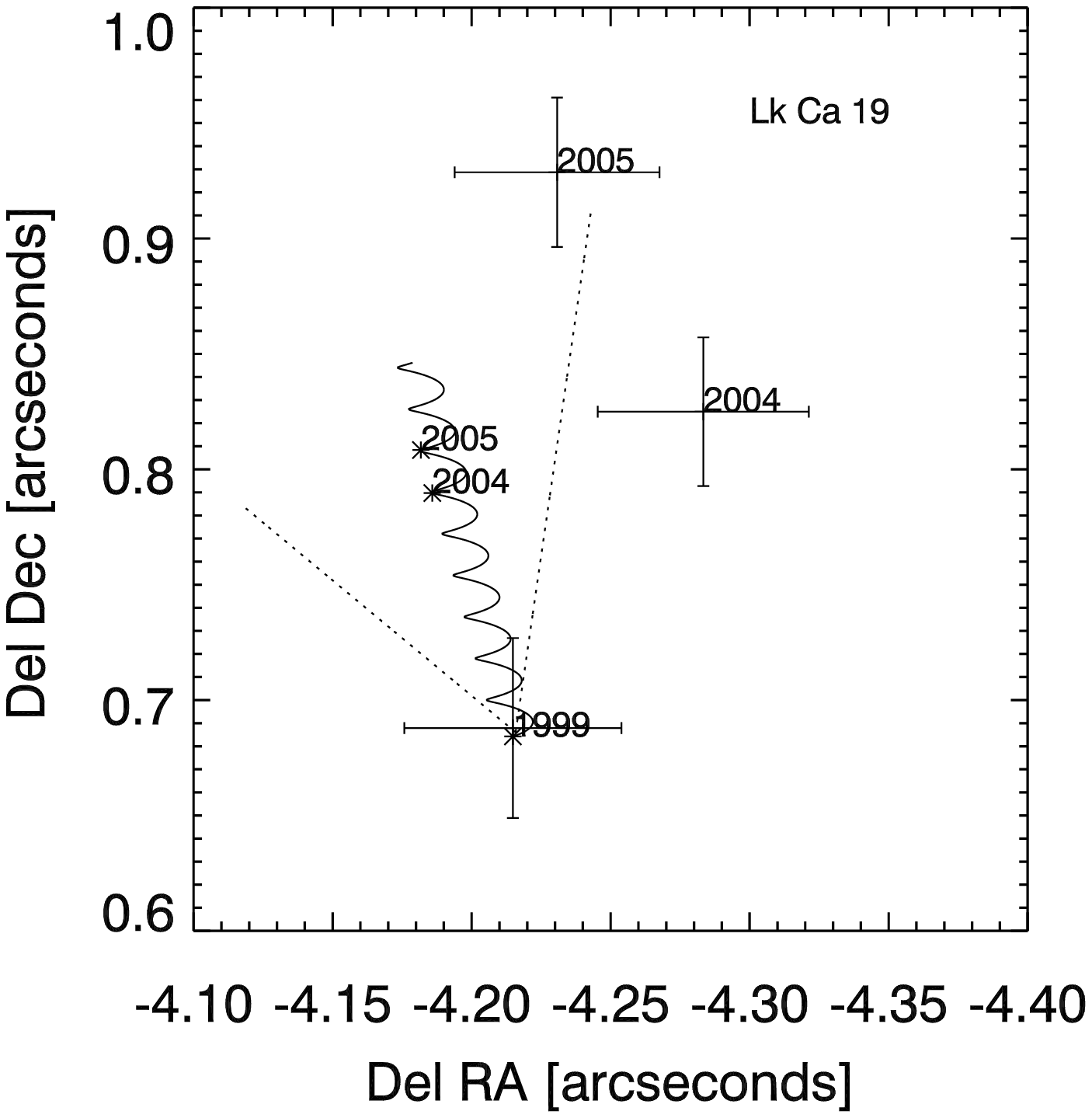}{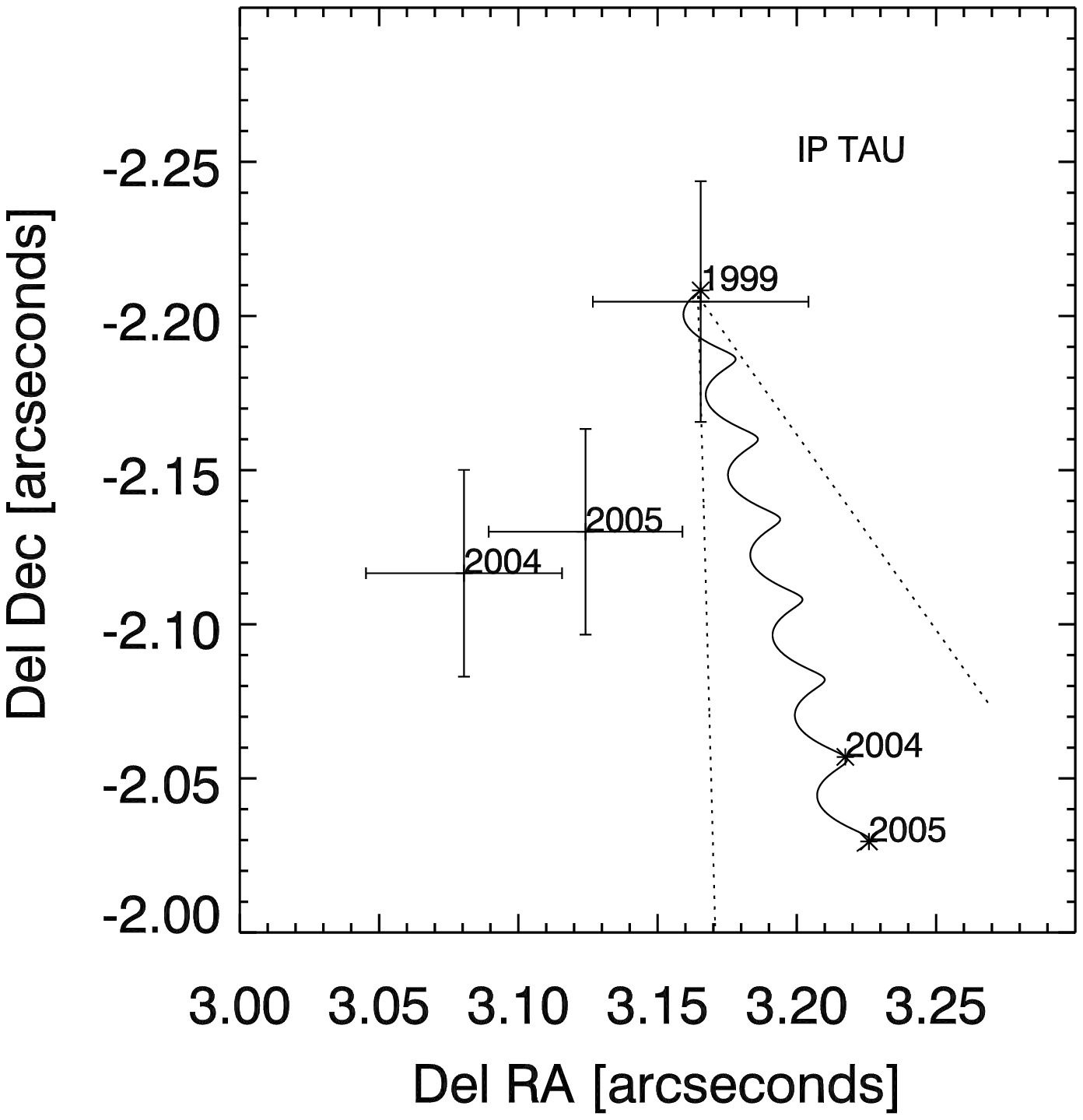}
\figcaption{Plot of the offset in RA and Dec between BP Tau, GK Tau, IP Tau, and LkCa19
and their companion candidates.
The WFPC 2 data point taken in Jan 1999 is used as the initial data point. Each measured
offset is noted with a cross and an epoch label. The curvy solid line shows the
expected motion of the star assuming measured proper motions from Frink et al. (1997). 
The expected offset of the companion if it
were a steady background object is labeled on the proper motion curve with epoch
values (2004,2005). Table~\ref{chitab} lists the reduced $\chi^2$ values associated
with fits to the data points which assume the companion candidates has
common and non-common proper motion with their primary stars. \label{pmplot}}
\end{figure}

Figure~\ref{pmplot} plots the offset in RA and Dec of companion candidates 
to BP Tau, IP Tau, GK Tau, and LkCa 19. These four sources
all have objects within at least 4.5 arcseconds (675 AU). 
While the probability of these companion candidates being background sources
goes up with their separation from the target star, the discovery of a
number of brown dwarf companions at wide separations ($>$200 AU) makes
these companion candidates worth investigating further. 
The crosses denote the positional offsets at the observing epochs of the Palomar data (2004, 2005)
as well as data taken from Hubble Space Telescope 
(HST) WFPC2 and NICMOS data when available. The HST data was collected in 1999 as part
of a program to detect faint debris disks (Krist et al. 2000) and therefore provides
a long time baseline for the determination of common proper motion. In most cases, the
target star in the WFPC2 data is saturated and the position of the star is estimated
from the intersection of the diffraction spikes. 
The size of the 
crosses denote the 3$\sigma$ positional uncertainties (1$\sigma$$\sim$12 mas for the
Palomar data and 1$\sigma$$\sim$3 mas for the WFPC2 data, J. Krist, private comm.)
The curvy solid
lines depict the changes in the offsets expected if the companion were a stationary, 
background object. The dotted lines represent the errors in the 
published proper motions (Frink et al. 1997). 
If the companions are
associated with the target stars, the offsets would coincide 
with one another at all epochs since the objects would share the same space motions. 
Table~\ref{chitab} lists the reduced chi-squared values estimated from the
positional data and uncertainties. Two hypotheses are tested - common proper motion
and non-common proper motion. In the first scenario the $\chi^2$ is derived from
the assumption that all the data points should lie on top of the first
epoch (1999) data point. For the second scenario, the $\chi^2$ is derived
from the assumption that the data points should lie on the vector produced
by the change in the offset between the science target and stationary background
star. The uncertainties for this case include both the positional uncertainties and
the uncertainty in the proper motion of the T Tauri star (Frink et al. 1997). 
An unmodeled source of uncertainty that may inflate the $\chi^2$s in the second scenario  
is the unknown proper motion of the companions. Based on this analysis, 
we conclude that there is evidence for common proper motion for the 
companion to GK Tau, evidence for
non-common proper motion for the companion to BP Tau and ambiguous evidence
for either scenario for IP Tau and LkCa19. These last two sources would benefit
from more accurate positional data and a longer time baseline. If the GK Tau 
companion is truly a physical companion, then based on its K$_s$ 
magnitude it would be roughly a M2-3 star with mass $\sim$0.5 M$_\sun$.

\begin{deluxetable}{lcc}
\tablecaption{Reduced $\chi^2$ Values from Common and Non-common proper motion fits\label{chitab}}
\tablehead{ \colhead{Target}  & \colhead{Reduced $\chi^2$$_{Common PM}$} & \colhead{Reduced $\chi^2$$_{Non-common PM}$} \\
}
\startdata
BP Tau  &  26  & 2.5 \\
GK Tau  &  1.4 & 4.1 \\
IP Tau  &  4.2 & 4.4 \\
LkCa19  &  43  & 6.3 \\
\enddata
\end{deluxetable}

We have begun a similar high contrast imaging survey of the SIM-YSO targets in
the Upper Sco subgroup of the Sco Cen association (1-2 Myr, 145 pc) using the NACO 
camera on the VLT. The NACO camera has been used to discover a number of 
low-mass companions including 2M1207 and GQ Lupi 
(Neuh{\"a}user et al. 2005; Chauvin et al. 2005). It has a pixel scale
of 27 mas pixel$^{-1}$, a FOV of 28$''$ with the potential of 
achieving sensitivities of $\Delta$K$_s$ of 10 at 1$''$ (Chauvin et al. 2004). 
So far, we have collected data on 20 targets with a number of them having companion 
candidates within one arcsecond. We will collect second epoch observations for
these sources at a later date. 

\begin{deluxetable}{lcccccc}
\tablecaption{Palomar Imaging Sensitivities \label{sentab}}
\tablehead{ \colhead{Target} & \colhead{T$_{int}$ [sec]} & \colhead{0$\farcs$5} & \colhead{1"} & \colhead{2"} & \colhead{5"} & \colhead{9"}
}
\startdata
HII 1032 & 1200 & 12.84 & 13.55 & 15.91 & 18.01 & 18.17 \\
HII 1136 & 1200 & 14.26 & 14.99 & 16.74 & 17.86 & 17.65 \\
HII 1797 & 1200 & 12.17 & 12.04 & 14.53 & 17.19 & 17.28 \\
HII 1794 & 1200 & 12.18 & 12.53 & 14.70 & 17.78 & 17.87  \\
HII 489  & 1200 & 14.68 & 13.93 & 16.20 & 17.57 & 17.76 \\
\hline  & &  &  &  &  &  \\
AA Tau  & 1200 & 13.31 & 11.44 & 15.20 & 17.16 & 17.36  \\
BP Tau  & 1200 & 11.08 & 11.65 & 14.26 & 17.41 & 17.57  \\
CI Tau & 1200 &  12.02 & 12.10 & 13.68&  17.67 & 17.67  \\
DL Tau  & 1200 & 12.51 & 13.17 &15.24 & 17.22 & 17.20   \\
DN Tau   & 900 & 12.80 & 10.91 & 14.13 & 17.13 & 17.18  \\
DQ Tau  & 1200 & 12.84 & 13.24 & 15.01 & 17.49 & 17.39  \\                
GK Tau  & 1200 & 11.58 & 11.22 & 12.90 & 16.94 & 17.15  \\
GI Tau & 1200 & 12.00 & 12.25 & 14.16 & 17.26 & 17.34  \\
IP Tau  & 1200 & 13.86 & 13.00 & 15.78 & 17.04 & 17.11  \\
IQ Tau & 1200 &  12.27 & 13.18 & 14.93 & 16.82&  16.93  \\
LkCa 19 & 900 & 12.46 & 12.55 & 14.93 & 17.97 & 18.22  \\
V830 Tau & 1200 & 13.59 & 14.85 & 17.09 & 20.15 & 20.36  \\
\enddata
\end{deluxetable}

\subsection{Comparison to Other Surveys}

There have been a few other ground AO and space-based surveys for low-mass companions to 
young stars (1-200 Myr) which have found a few brown dwarfs with separations between 
75 to 1000 AU. These surveys had sample sizes ranging from 30-100 targets with two 
surveys finding 2-3 brown dwarfs when targeting stars in nearby associations 
(Metchev 2005; Lowrance et al. 2005) and one survey finding no brown dwarfs 
when targeting X-ray selected T-Tauri stars in the Chamaeleon and Sco-Cen OB 
associations (Brandner et al. 2000). Metchev (2005) has estimated a completeness
corrected percentage of brown dwarfs at $\sim$7$\pm$3\% (1$\sigma$ confidence) 
around F5-K5 stars with an age and separation range
of 3-500 Myr and 30-1600 AU. Therefore our finding no 
very low-mass objects in a survey of 30 stars is consistent with previous 
programs given the stated uncertainties. Whether there is a significant 
dearth of brown dwarfs or planetary mass objects around young stars still 
requires a larger sample of targets with similar sensitivities to allow for a 
direct comparison of detection statistics. 

Stellar multiplicity surveys of star formation regions such as
Taurus and the Pleiades have revealed binary companion fractions 
of 60\% (over 20-500 AU) and 30\% (over 1-900 AU), 
respectively (Ghez et al. 1993; Bouvier et al. 1997). However, we
used these and other surveys mentioned previously (see \S 2) to remove 
stars with stellar companions within two arcseconds in defining our initial
sample. Thus it is not surprising that our survey did not add to the statistics for stellar multiplicity in these clusters. 

\subsection{Speckle Imaging and Additional AO imaging of the Taurus Sample}

To complement the adaptive optics survey at Palomar, a number
of the targets in the sample have been observed using high
resolution imaging techniques at the Keck observatory.  The
purpose of this project was to look for companions at separations too
close to be resolved with Palomar, and too wide to be
detected via spectroscopic techniques.  A summary of the
results of this portion of the project is given in Table~\ref{specktab}.

In total, seventeen objects in Taurus, TW
Hydrae, or AB Doradus were surveyed at Keck.  Three of these
were imaged using speckle interferometry at K-band (2.2 $\micron$) on Keck I,
and fourteen were observed using adaptive optics and K' (2.3 $\micron$) or
L'-band (3.5 $\micron$) on Keck II.  The dates of these observations are given
in Table~\ref{specktab}, along with the total exposure time on each
target.  For details on the data reduction and analysis of
both the speckle and the AO data taken at Keck, please see
Konopacky et al (2007).

For all 17 sources, no companions were found to within
0$\farcs$05.  For each source we estimated our
sensitivity to companions by finding the limiting detectable flux ratio
with respect to the source as a function of radius, and then
using the models of Baraffe et al. (1998; $\alpha$ = 1.0) to
convert these flux ratio limits into mass limits for the closest radius bin of 0$\farcs$5. 
In general, the speckle
measurements probe regions closer than 0$\farcs$05 with much
greater sensitivity than AO, but given the combination of the
two techniques for this survey, we cut off our official
completeness at 0$\farcs$5. We plan to observe the remaining targets
visible from the Northern hemisphere with either Keck AO or speckle imaging 
and are beginning a 
survey of the targets in the Southern hemisphere using Lucky imaging, a 
similar observational technique at optical wavelengths (Law et al. 2006).

\section{Interferometric Observations of Stars in Taurus and the Pleiades} 

The Keck Interferometer (KI) was used to make near-infrared, long-baseline
interferometry observations of three sources in the Pleiades and five
sources in Taurus.  These observations are part of a long-term
program to study the multiplicity fraction of these sources.
In particular, the interferometer is sensitive to companions
within 30 mas of the primary star and with a K magnitude
difference of 3 magnitudes or less.  

The observations were taken on 10 November, 2006 with KI configured in
the 5-channel, K band (2.18 microns central wavelength) mode described
in Colavita et al. (2003).  Calibrators were chosen to match the targets
in K magnitude and were reduced with the standard parameters (Colavita
et al 2003), including the ratio correction for imbalanced flux on the
two paths of the interferometer.  The calibrator sizes were all set to
0.1 $\pm$ 0.05 mas diameter, but in all cases, the calibrator
diameter did not contribute significantly to the final uncertainty.
The primary KI data product is the normalized visibility amplitude
squared, for which a value of 1 indicates an unresolved source.  The
uncertainty given is the quadrature combination of the scatter in the
target measurement and the uncertainty in the system visibility (which
is this case is dominated by the scatter in the calibrator
measurements).  Table~\ref{kecktab} lists the targets, the numbers of integrations,
the calibrated visibility and total uncertainty and the calibrators
used for each target.  The uncertainty varies due to the number of
integrations and the difficulty in observing those sources with
fainter V magnitudes.

All targets observed in this sample, except V830 Tau, are unresolved; i.e. consistent
with a point source at this resolution.  At the distance
to Taurus or the Pleiades the central star will be unresolved
($<$0.1 mas). Using the best uncertainties of 0.06, 
we can place limits on the size of the emission of 1.3 mas (1.8 AU) in
diameter (3 $\sigma$) for a uniform disk
or on the presence of over-resolved (diffuse) emission
within the 50 mas field-of-view of $<$10\% (3 $\sigma$).
Additional observational epochs are needed to constrain the
multiplicity. V830 Tau has visibilities estimated from earlier epochs
of data (2003, 2004, Akeson et al. 2005) which differ from the expected value for
an unresolved source by a few sigma (see Table~\ref{kecktab}). Future observations
of this star with KI are planned to determine the nature of the resolved emission.

\section{Radial Velocity Vetting}

Some of the target stars will have stellar or sub-stellar companions that are not 
detectable by direct imaging (separations $<$50 AU). We are conducting a number of 
radial velocity (RV) surveys of potential SIM-YSO stars to determine whether the targets 
have unseen companions that might complicate the astrometric detection of planetary-mass 
objects. For  example,  a 20 M$_J$ companion in a 1 year orbit 
around a 0.8 M$_\sun$ star has an RV amplitude of 660 m s$^{-1}$. At 140 pc, this object
would produce an astrometric signature of 160 $\mu$as that would swamp the signal from 
any lower-mass planets. Located just a few milli-arcseconds from its parent star, 
the brown dwarf companion would be undetectable by direct, non-interferometric imaging. The goal of 
the RV program is to achieve accuracies on the order of $<$500 m s$^{-1}$ over 
3-4 years depending on the $v\, sin(i)$ and photospheric instabilities of each 
star with the goal of getting limits on stellar and sub-stellar companions on orbits interior 
of 5 AU. 

\begin{figure}[ht]
\plotone{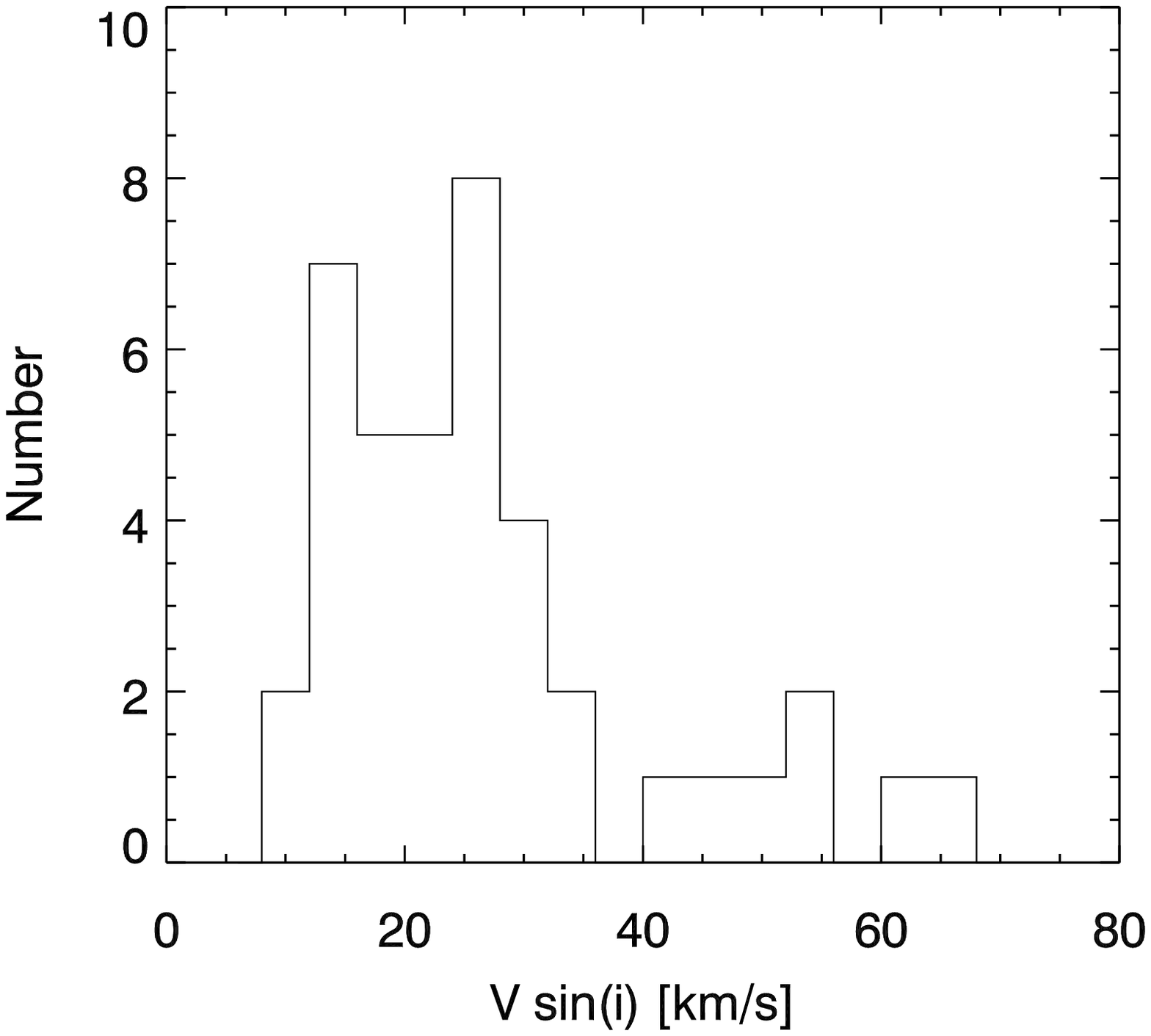} 
\figcaption{Histograms of the values of 
{\it v sin(i)} estimated from the Sco Cen sample.  \label{vsinihist}}
\end{figure}

There are two radial velocity surveys presently being conducted on the SIM-YSO 
targets in both the Northern and Southern hemispheres. The Southern survey was
begun in July 2003 using the echelle spectrograph (R=25,000) on the CTIO 4-m 
telescope (S. Mohanty, PI). All 42 of the targets in this survey were in the 
Sco-Cen association. Only one epoch of these  targets have been collected to 
date because of the unfortunate retirement of this  instrument on this telescope. 
Many of the stars in Sco-Cen turned out to be fast rotators 
($>$10 km s$^{-1}$, see Table~\ref{vsinitab} and Figure~\ref{vsinihist}) which is not surprising
since, at 10-20 Myrs of age, there is ample time for the stars to spin-up due to stellar
contraction.

The Northern survey began in October 2004, and used the
coude echelle spectrograph (R=60,000) on the 2.7-m
Harlan J. Smith telescope at McDonald Observatory
(L. Prato, PI).  Survey targets are located in the Pleiades
cluster and in the Taurus star forming region.   To date, 51
objects have been observed at multiple epochs with an RV
precision of 140 m s$^{-1}$.  This survey makes use of simultaneous
BVR photometry to search for correlations between
rotation and RV periods, indicative of star spot modulation
rather than the presence of a low-mass companion.
Initial results which include a few targets with potential companions or signs of starspots
will be presented in a forthcoming publication (Huerta et al. 2007; Huerta et al. 2005). 

\begin{deluxetable}{lclc}
\tablecaption{{\it V sin(i)} values for Sco-Cen targets \label{vsinitab}}
\tablehead{ \colhead{Target} & \colhead{{\it Vsin(i)}} & \colhead{Target} & \colhead{{\it Vsin(i)}} \\
            \colhead{}       & \colhead{[km s$^{-1}$]}  & \colhead{}       & \colhead{[km s$^{-1}$]} 
}
\startdata
TYC8648-446-1 & 10 & TYC8238-1462-1 & 25 \\
TYC8283-264-1 & 10 & TYC8654-1115-1 & 25 \\
TYC8282-516-1 & 13 & TYC8655-149-1 & 25 \\
TYC8645-1339-1 & 15 & TYC9245-617-1 & 25 \\
TYC9246-971-1 & 15 & TYC7833-2559-1 & 25 \\
TYC7796-1788-1 & 15 & TYC8295-1530-1 & 25 \\
TYC9244-814-1 & 15 & TYC7353-2640-1 & 25 \\
TYC7319-749-1 & 15 & TYC8636-2515-1 & 30 \\
TYC8317-551-1 & 15 & TYC8646-166-1 & 30 \\
TYC8242-1324-1 & 18 & TYC7310-2431-1 & 30 \\
TYC8249-52-1 & 18 & TYC8294-2230-1 & 30 \\
TYC8259-689-1 & 18 & TYC7824-1291-1 & 33 \\
TYC8297-1613-1 & 18 & TYC8652-1791-1 & 35 \\
TYC7848-1659-1 & 18 & TYC7852-51-1 & 40 \\
TYC8640-2515-1 & 20 & TYC8633-508-1 & 45 \\
TYC8644-340-1 & 20 & TYC8248-539-1 & 50 \\
TYC7811-2909-1 & 20 & TYC7333-1260-1 & 55 \\
TYC8667-283-1 & 23 & TYC7851-1-1 & 55 \\
TYC7822-158-1 & 23 & TYC7783-1908-1 & 63 \\
TYC9212-2011-1 & 25 & TYC8270-2015-1 & 65 \\
\enddata
\end{deluxetable}

\section{Photometric Monitoring}

The photospheric activity that affects radial velocity and transit measurements 
affects astrometric measurements as well, but, as we will now show, at levels 
consistent with the secure detection of gas giant planets with SIM. From measurements of 
photometric variability (Herbst et al. 1994; Bouvier \& Bertout, 1989; Bouvier et al. 1995) plus 
Doppler imaging (Strassmeier \& Rice 1998), T Tauri stars are known to have 
active photospheres  with large starspots covering significant portions of 
their surfaces (Schussler et al. 1996), as well as hot spots due to infalling, 
accreting material  (Mekkaden 1998). Day-to-day changes arise primarily because 
of rotation whereas month-to-month variations reflect changes in the spot 
sizes and their  distribution across the surface. Long term monitoring is 
essential because different stars  have different levels of magnetic activity, 
and these levels can change with time. These effects can produce large 
photometric variations which can  significantly shift the photocenter of a star. 
In the simplest  approximation, a completely black starspot, covering a small 
fraction, $\beta<<1$, of a  stellar hemisphere, will shift the photocenter by an angle

$$\Delta\phi(\theta) \sim \beta {\rm sin\theta \, cos\theta } { {R_\star} \over 
{D_\star}} = 33.4 \, \beta \, { {R_\star} \over {R_\odot}} { {140\, pc} \over 
{D_{pc}}} {\rm sin\theta\, cos\theta }\,\,\, [\mu{\rm as}] \,\,\, (2) $$

\noindent where R$_\star$ is the stellar radius, D$_*$ is the distance
to the star, and $\theta$ is the longitude of 
starspot relative to the line of sight. We have assumed that the spot is on the 
star's equator and that the star is observed edge-on relative to its rotation axis. 
The shift in the photocenter, $\Delta\phi(\theta)$, will increase as the spot rotates  
away from a face-on longitude ($\propto$ sin$\theta$), following the rotation of the  
star as the spot shrinks in projected area ($\propto$ cos$\theta$) and eventually goes  
behind the star. Relative to the fractional change in stellar brightness,  
$\Delta I/I(\theta)=\beta\, {\rm cos}\theta$ and averaging over $-\pi/2<\theta<\pi/2$,  
we get a root mean square (rms) dispersion in the location of the photocenter  
(the ``astrometric jitter'') given by,

$$<\Delta\phi>= \alpha < { {\Delta I} \over {I}} >  { {R_\star} \over {D_\star}} = 
1.8 { {R_\star} \over {R_\odot}} { {140 \, pc} \over {D_{pc}}} { {\Delta {\rm R 
(mag)}}\over{0.05 \, {\rm mag}}} \,\, [\mu{\rm as}] \,\,\,\, (3) $$

\noindent where $\alpha$ is a geometric term of order 1.1. A more careful analysis 
takes into account the fact that the spots are not completely black, but rather  
emit with a temperature of $\sim$1,000 K cooler than the photosphere, are located 
over a range of typically high latitudes, and that an ensemble of stars will be observed at 
 random angles to the line of sight. A Monte Carlo simulation  shows a linear 
relationship like that of Eqn (3), but with a smaller coefficient:

$$<\Delta\phi (Monte \,\,Carlo)>= 0.9 \, { {R_\star} \over {R_\odot}} 
{ {140 pc} \over {D_{pc}}} { {\Delta {\rm R (mag) }} \over {0.05 \, {\rm mag}}} 
\,\,\, [\mu{\rm as}] \,\,\,\, (4).$$

For a typical T Tauri star radius of 3 $R_\odot$ in Taurus, we see that the astrometric 
jitter is less than 3 $\mu$as for R-band variability less than or equal to 0.05 mag 
(1$\sigma$). Thus,  the search for Jovian planets with astrometric amplitudes 
greater than 6 $\mu$as is possible for stars less variable than 
about 0.05 mag in the visible even without a correction for jitter that may be 
possible using astrometric information at multiple wavelengths. Other astrophysical 
noise sources, such as offsets induced by the presence of nebulosity and stellar 
motions due to disk induced non-axisymmetric forces are negligible for appropriately 
selected stars. Finally, it is worth noting that searching for terrestrial planets 
will be difficult until stars reach an age such that their photometric variability falls
well below 0.01 mag and the corresponding astrometric jitter
below 1 $\mu$as. Even then, color dependent astrometric
corrections may be needed for the most sensitive measurements. 

\begin{figure}[ht]
\plotone{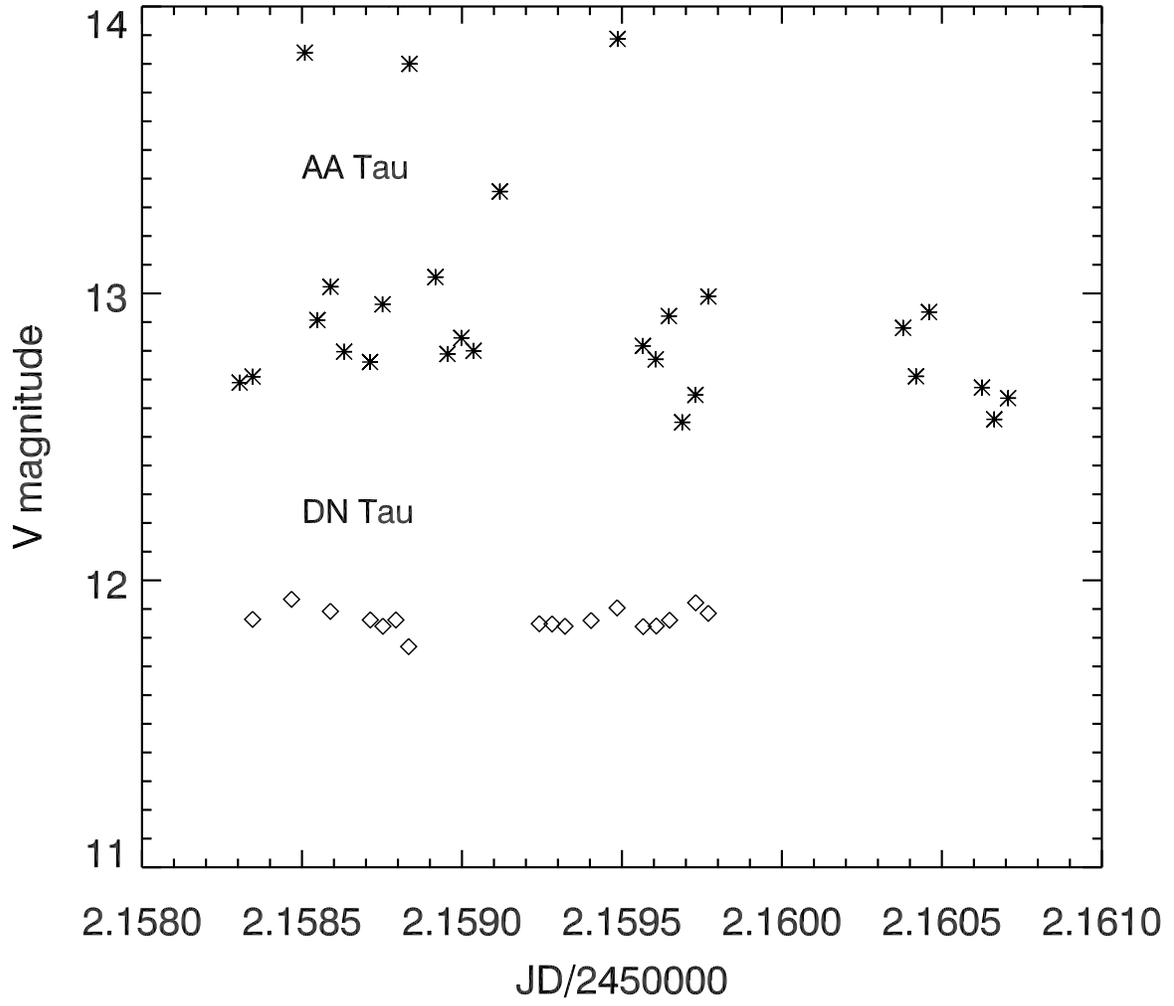} 
\figcaption{ Plot of the V band photometry taken for AA Tau and DN Tau. The standard deviations of the photometry for these two 
sources is 1.5 and 0.17 mag, respectively, making the first source a problematic
SIM-YSO target\label{varplot}}
\end{figure}

Since young stars have active photospheres it is important that we asses the 
degree of photospheric activity to determine the best targets for the
program. The SIM-YSO team has two separate programs conducting photometric 
monitoring  of the sample targets in the Northern and Southern hemispheres. 
The Southern targets are being observed in R band (0.9 $\micron$) at CTIO using the SMARTS 
(Small and Medium Aperture 
Research Telescope System) program (M. Simon, PI) which is comprised of 
small (0.9-1.5 meter) telescopes in the Southern  Hemisphere.  To date, they have observed 132 stars in the Sco-Cen  and Upper-Sco 
associations. Figure~\ref{varplot} plots the R magnitudes of AA Tau and DN Tau taken 
with the SMARTS survey. The standard deviations of the photometry for these two 
sources is 0.17 and 1.5, respectively, making the latter source a problematic
SIM-YSO target.

The Northern component of the SIM-YSO sample which includes stars in Taurus and 
the Pleiades has been monitored photometrically with small telescopes at the 
Maidanak Observatory (W. Herbst, PI; Grankin et al. 2007). Approximately 
ten data points are obtained  on each star during each season to sample the range of 
the variability. The data are taken primarily in the B, V, and R bands, with a small 
amount taken in the U band.   Forty-two stars are on the program and about 450 
individual measurements have been obtained each season.

Out of those targets observed to date in both the Northern and
Southern samples, 33\% of them have photometric variability
that produces astrometric noise greater than 3 $\mu$as 
(or roughly 1$\sigma\sim$ 0.05 mag) in 
either the V or R band (see Table~\ref{phottab}). 
 Simultaneous monitoring of
the variable stars both photometrically and with radial velocity measurements during
the SIM observations might allow us to model the jitter and derive 
accurate astrometry.
Those targets exhibiting 
significant amounts of photometric variability will be monitored further
to asses whether they should remain in the target list or regulated to Wide
Angle observations which will be sensitive to Jupiter mass planets further  
than one AU from the star.

\section{Discussion}

Through a series of precursor programs to observe all of the stars in
the SIM-YSO target list it becomes clear that the observable affecting
the viability of the targets the most is photometric variability. 
To date, 22\% of the stars (33\% of those stars observed) in the target list 
have variability which contributes
more than 3 $\mu$as of noise to the astrometric measurements. 
Table~\ref{samtab}, which provides basic information on the entire
SIM-YSO list, has been separated into ``low variability", ``high variability"
and ``not observed" sections based on the degree of their variability and whether they
have been observed to date.  Each sublist is then ranked by the astrometric signal expected 
for a Jupiter mass planet at 1 AU from the stars which is listed under ``Signal" in 
the table. To date, 67\% of the sample has variability data with the 
remainder expected to be completed within the next year or two. 

To replace those stars which might be lost to photometric variability, we will examine 
the literature for objects in other clusters or for newly classified T-Tauri stars. 
While many new young stars are being discovered in relatively nearby moving groups 
(i.e. Song et al. 2003), these typically have ages of 10-25 Myr. These stars can  be  
observed inexpensively  with SIM's Wide Angle mode and will probably not have 
problems of excessive variability. We will concentrate on 
finding bright enough replacement stars in the 
1-5 Myr age range to avoid skewing our sample toward older stars. 

\subsection{Observing Scenarios and Reference Stars}

While the nearer target stars  (d$<$50 pc) will have planetary astrometric signatures 
large enough to be detected by Wide Angle observations, 
more distant or more massive stars will have astrometric 
signatures on the order of 6 $\mu$as requiring Narrow Angle observations. 
Narrow Angle observations require 
a set of at least three reference stars within the $\sim$1$^o$5 field-of-view used
for these observations. These
reference stars must themselves be astrometrically stable to within $<$ 4 $\mu$as. The
best reference stars are K giants at distances $>$500 pc.  There is an additional 
precursor program presently making photometric observations of the pool of 
reference stars available for every SIM-YSO 
target (or clusters of targets). When choosing the reference stars, we used a combined 2MASS-Tycho 2 catalog to  select K giants 
based on visible-near IR colors as well as from reduced proper motions (RPM). The 
following selection criteria were also used to make the initial lists of reference stars: 1) separation 
from target $<$ 1.25 degrees, 2) 0.5 $<$ (J-K$_s$) $<$ 1.0, 3) 1.0 $<$ (B$_{Tycho}$-V$_{Tycho}$) 
$<$ 1.5, 4) V$_{Tycho}$ $<$ 10, and 5) RPM = K$_s$ + 5 log($\mu$) $<$ 1. While some of the
 reference stars have published spectral types,  
many stars do not have any spectral type at all.  We have begun a program of
verification of the luminosity classes of the photometrically selected sample
using the SMARTS telescopes. 

Observing the stars in Narrow Angle mode requires roughly five times more integration 
time than doing so in Wide Angle mode, primarily due to the necessity of observing 
3-5 reference stars. Putting together the final program will require a balance 
between observing fainter, more distant and young objects in Narrow Angle mode 
versus the brighter, closer, and older stars in Wide Angle mode.

\section{Conclusions}

We have presented the results for a number of precursor programs
aimed at creating a robust list of young stars to be observed
as part of a Key project for the SIM PlanetQuest mission. This program will detect 
Jupiter-mass planets at distances of $\sim$1-5 AU from the star, thereby probing 
planet formation at distances comparable to where radial velocity planets are 
found around mature, main-sequence stars. The imaging surveys did not find any stars with bright, nearby companions that
could pose a problem for SIM; although we did find
a potential M star companion 2$^{\prime\prime}$.4 away from GK Tau.  The radial velocity 
surveys may have found one or two stars with close-in companions. 
In the near future, the RV surveys will be supplemented with 
high spectral resolution RV surveys in the near-infrared. These observations
are not as affected by the photometric variability 
of the star and are expected to achieve RMS accuracies of down to 100 m $^{-1}$ thereby allowing 
for the detection of lower mass objects. One selection effect we will
have to guard against is losing too many of the youngest stars due to large
photometric variations. We will continue to supplement our target list  with 
additional stars which meet our basic  criteria as well as investigate ways to mitigate 
astrometic jitter using multi-wavelength data from SIM itself.

\begin{acknowledgements}
LP and MH thank our colleagues C. Johns-Krull, P. Hartigan,
and D. Jaffe for their collaboration on the McDonald Observatory
project.
Based on observations obtained at the Hale Telescope, Palomar Observatory,  
as part of a continuing collaboration between the California Institute of 
Technology, NASA/JPL, and Cornell University. The research described in this 
publication was carried out at the Jet Propulsion Laboratory,
California Institute of Technology, under a contract with the National 
Aeronautics and Space Administration. This publication makes use of data products 
from the Two Micron All Sky Survey, which is a joint project of the University of 
Massachusetts and the Infrared Processing and Analysis Center/California Institute 
of Technology, funded by the National Aeronautics and Space Administration and 
the National Science Foundation.
\end{acknowledgements}

\begin{center}
\begin{deluxetable}{lcccccccc}
\tablecolumns{9}
\tablewidth{0pt}
\tabletypesize{\footnotesize}
\tablecaption{Keck Imaging Sensitivities \label{specktab}} 
\tablehead{
\colhead{Object} & \colhead{Date} &\colhead{Method} &
\colhead{Total Exp.} & \colhead{$\Delta$K Lim} & \colhead{$\Delta$K Lim} &
\colhead{$\Delta$K Lim} & \colhead{Mass Lim}  & \colhead{Band}\\
\colhead{ } & \colhead{Obs} & \colhead{(Sp or AO)} &
\colhead{Int. Time [sec]}&\colhead{0$\farcs$05} &
\colhead{0$\farcs$1} & \colhead{$\geq$0$\farcs$5} & \colhead{0$\farcs$05 [M$_{\sun}$]} & \colhead{}
}
\startdata
TWA 23 & 2005 May 27 & Sp &78.1 & 2.9 & 3.7 & 6.9 & 0.10 & K \\ 
TYC7660-0283 & 2005 May 27 & Sp &78.1 & 3.9 & 4.1 & 6.5 & 0.08 & K \\
GM Tau & 1997 Oct 12 & Sp & 156.2 & 3.9 & 3.9 & 6.5 & $\lesssim$0.02& K \\
Anon 1 & 2005 Dec 12 & AO & 120 & 3.8 & 4.4 & 6.2 & 0.06& K \\
BP Tau & 2005 Dec 12 & AO & 120 & 5.5 & 4.3 & 6.2 & 0.03 & K \\
DG Tau & 2005 Dec 12 & AO & 120 & 5.8 & 3.5 & 5.9 & 0.02 & K \\
HD 283572 & 2005 Dec 12 & AO & 54.3 & 4.6 & 4.8 & 6.3 & 0.03 & K \\
IP Tau & 2005 Dec 12 & AO & 120 &3.7 & 4.5 & 5.3 & 0.06 & K \\
IQ Tau & 2005 Dec 12 & AO & 120 & 5.0 & 4.7 & 5.9 & 0.03 & K \\
V1072 Tau & 2005 Dec 12 & AO & 120 & 5.1 & 5.5 & 6.5 & 0.03 & K \\
DN Tau & 2005 Dec 12 & AO & 120 & 4.7 & 5.1 & 6.2 & 0.03 & K \\
V830 Tau & 2005 Dec 12 & AO & 60 & 2.9 & 3.9 & 5.8 & 0.15 & K  \\
DR Tau & 2005 Dec 12 & AO & 54.3 & 3.4 & 3.9 & 6.6 &  0.14 & K \\
HIP 113579 & 2005 Jul 16 & AO & 3.0 & 3.7 & 3.5 & 5.7 & 0.24 & L \\
HIP 113597 & 2005 Jul 16 & AO & 6.0 & 3.0 & 5.1 & 5.7 & 0.12 & L \\
HIP 114066 & 2005 Jul 16 & AO & 6.0 & 2.9 & 3.0 & 5.3 & 0.09 & L \\
HIP 115162 & 2005 Jul 16 & AO & 6.0 & 3.0 & 4.9 & 5.2 & 0.39 & L \\
\enddata
\end{deluxetable}
\end{center}

\begin{deluxetable}{lccccc}
\tablecaption{Keck Interferometry Results \label{kecktab}}
\tablehead{ \colhead{Target} & \colhead{K mag} & \colhead{\# ints} & \colhead{avg V$^2$} & \colhead{avg Sig} & \colhead{Calibrators} 
}
\startdata
DN Tau     & 8.0 & 1 &1.02 &0.09 &HD283444,HD283886 \\
V830 Tau$^a$   & 8.4 & 1 &0.85 &0.10 &HD283668,HD282230,HD29334\\
V830 Tau$^b$   & 8.4 & 1 &0.89 &0.07 &HD283668,HD282230,HD29334\\
V830 Tau   & 8.4 & 1 &1.08 &0.13 &HD283444,HD283886\\
V1171 Tau  & 9.2 & 2 &0.98 &0.05 &HD24132,HD23289,HD284316\\
V1072 Tau  & 8.3 & 1 &1.13 &0.10 &HIP19757,HD285816\\
V1075 Tau  & 8.9 & 1 &1.08 &0.10 &HIP19757,HD285816\\
HD 282973  & 8.6 & 2 &0.97 &0.06 &HD24132,HD23289,HD284316\\
HD 282971  & 8.7 & 2 &0.99 &0.06 &HD24132,HD23289,HD284316\\
HD 23584   & 8.3 & 2 &0.98 &0.06 &HD24132,HD23289,HD284316\\
\enddata
\tablenotetext{a}{Observation from Oct. 16 2003}
\tablenotetext{b}{Observation from Jan. 07 2004}
\end{deluxetable}

\begin{deluxetable}{lccccccccccc}
\tabletypesize{\scriptsize}
\tablecaption{SIM-YSO Sample \label{samtab}}
\tablehead{
\colhead{Name} & \colhead{Cluster} & \colhead{Distance} & \colhead{Spec} & \colhead{Age}   & \colhead{Star Mass$^a$}     & \colhead{Signal}   & \colhead{T-Tauri} & \colhead{V}    & \colhead{2MASS K$_s$} \\
               &                   & \colhead{[pc]}     & \colhead{Type}     & \colhead{[Myr]} & \colhead{[M$_\sun$]}    & \colhead{[$\mu$as]}& \colhead{Class}   & \colhead{[mag]}& \colhead{[mag]} 
}
\startdata
{\bf Low variability targets} &  &  &  &  &  &  &  &  &  \\
PreibZinn9964$^b$ &  USco  & 145 & M1 & 0.1 & 0.3 & 23.68 & ... &  & 8.36 \\
51Eri &  Beta Pic  & 29.8 & F0V & 20 & 1.5 & 21.51 & ... & 5.22 & 4.54 \\
PreibZinn9914 &  USco  & 145 & M0.5 & 0.5 & 0.3 & 19.50 & … & … & 9.09 \\
RECX10 &  Eta Cha  & 100 & K6 & 7 & 0.6 & 17.48 & WTT & 12.53 & 8.73 \\
PreibZinn9928 &  USco  & 145 & M0 & 1 & 0.4 & 15.79 & … & … & 8.80 \\
PreibZinn9913 &  USco  & 145 & M0 & 1.2 & 0.4 & 15.07 & ... &  & 8.88 \\
PreibZinn9919 &  USco  & 145 & K6 & 0.7 & 0.5 & 12.51 & ... &  & 8.11 \\
PreibZinn9974 &  USco  & 145 & K4 & 0.5 & 0.6 & 10.36 & ... &  & 8.46 \\
PreibZinn9936 &  USco  & 145 & K7 & 2.7 & 0.7 & 10.20 & … & … & 9.12 \\
PreibZinn9967 &  USco  & 145 & K5 & 1.1 & 0.7 & 10.05 & … & … & 8.56 \\
PreibZinn9969 &  USco  & 145 & K5 & 1.8 & 0.7 & 9.47 & ... &  & 8.56 \\
HII489 &  Pleiades  & 130 & F8 & 125 & ... & 9.25 & ... & 10.38 & 8.87 \\
HII1794 &  Pleiades  & 130 & F8 & 125 & ... & 9.25 & ... & 10.20 & 8.89 \\
HII2366 &  Pleiades  & 130 & G2 & 125 & ... & 9.25 & ... & 11.53 & 9.55 \\
TYC8283-2795-1 & Sco-Cen(UCL) & 130 &  & ... & ... & 9.25 & WTT & 10.79 & 8.98 \\
PreibZinn9950 &  USco  & 145 & K5 & 2.5 & 0.8 & 8.84 & … & … & 8.90 \\
L1551-55 &  Tau Aur  & 140 & K7 & 2 & ... & 8.59 & … & 13.22 & 9.31 \\
PreibZinn9939 &  USco  & 145 & K4 & 2.2 & 0.8 & 8.29 & ... &  & 8.73 \\
HII1124 &  Pleiades  & 130 & K3V & 125 & 1.0 & 7.79 & ... & 12.12 & 9.86 \\
PreibZinn9958 &  USco  & 145 & K2 & 1.2 & 0.9 & 7.62 & ... &  & 8.43 \\
HII1095 &  Pleiades  & 130 & K0V & 125 & 1.0 & 7.40 & ... & 11.92 & 9.67 \\
HII1309 &  Pleiades  & 130 & F6V & 125 & 1.0 & 7.40 & ... & 9.58 & 8.28 \\
HII1613 &  Pleiades  & 130 & F8V & 125 & 1.0 & 7.40 & ... & 9.87 & 8.57 \\
HII1797 &  Pleiades  & 130 & F9V & 125 & 1.0 & 7.40 & ... & 10.09 & 15.04 \\
HII1856 &  Pleiades  & 130 & F8V & 125 & 1.0 & 7.40 & ... & 10.20 & 8.66 \\
TYC8654-1115-1 & Sco-Cen(LCC) & 130 &  & 24 & 1.0 & 7.40 & WTT & 10.21 & 8.13 \\
TYC8295-1530-1 & Sco-Cen(UCL) & 130 & G5 & 21 & 1.0 & 7.40 & WTT & 10.98 & 8.90 \\
PreibZinn9945 &  USco  & 145 & K2 & 1.2 & 0.9 & 7.37 & ... & 11.17 & 8.04 \\
HD141569 &  none  & 35 & B9.5e & 5 & 4.0 & 6.87 & ... & 7.11 & 6.82 \\
TYC8667-283-1 & Sco-Cen(LCC) & 130 & G3/G5V & 23 & 1.1 & 6.72 & WTT & 9.31 & 7.62 \\
TYC7783-1908-1 & Sco-Cen(LCC) & 130 & G8IV: & 18 & 1.1 & 6.72 & WTT & 9.82 & 7.51 \\
TYC8258-1878-1 & Sco-Cen(LCC) & 130 &  & 15 & 1.1 & 6.72 & WTT & 10.62 & 8.27 \\
TYC9244-814-1 & Sco-Cen(LCC) & 130 & G3/G5III & 22 & 1.1 & 6.72 & WTT & 10.21 & 8.40 \\
TYC8270-2015-1 & Sco-Cen(UCL) & 130 &  & 17 & 1.1 & 6.72 & WTT & 10.91 & 8.69 \\
TYC7851-1-1 & Sco-Cen(UCL) & 130 & G9 & 17 & 1.1 & 6.72 & WTT & 10.63 & 8.36 \\
TYC7353-2640-1 & Sco-Cen(UCL) & 130 &  & 18 & 1.1 & 6.72 & WTT & 10.72 & 8.67 \\
CHXR8 &  Cham  & 140 & G0 & 100 & 1.1 & 6.24 & WTT & 11.45 & 9.73 \\
HII430 &  Pleiades  & 130 & G8V & 125 & 1.2 & 6.16 & ... & 11.40 & 9.47 \\
HII1032 &  Pleiades  & 130 & A2 & 125 & 1.2 & 6.16 & ... & 11.10 & 9.16 \\
HII1136 &  Pleiades  & 130 & G7V & 125 & 1.2 & 6.16 & ... & 12.02 & 12.14 \\
HII1275 &  Pleiades  & 130 & K0V & 125 & 1.2 & 6.16 & ... & 11.47 & 9.53 \\
TYC8646-166-1 & Sco-Cen(LCC) & 130 &  & 11 & 1.2 & 6.16 & WTT & 10.50 & 8.18 \\
TYC8636-2515-1 & Sco-Cen(LCC) & 130 &  & 11 & 1.2 & 6.16 & WTT & 10.58 & 8.12 \\
TYC8633-508-1 & Sco-Cen(LCC) & 130 & K2IV:+… & 16 & 1.2 & 6.16 & WTT & 9.41 & 7.65 \\
TYC9245-617-1 & Sco-Cen(LCC) & 130 &  & 10 & 1.2 & 6.16 & WTT & 10.01 & 7.55 \\
TYC8652-1791-1 & Sco-Cen(LCC) & 130 & F6/F7 & 16 & 1.2 & 6.16 & WTT & 10.35 & 8.48 \\
TYC8259-689-1 & Sco-Cen(LCC) & 130 &  & 14 & 1.2 & 6.16 & WTT & 10.48 & 8.10 \\
TYC8248-539-1 & Sco-Cen(LCC) & 130 & G1/G2 & 26 & 1.2 & 6.16 & WTT & 10.10 & 8.54 \\
HD120411 & Sco-Cen(UCL) & 130 & G1V & 20 & 1.2 & 6.16 & WTT? & 9.79 & 8.16 \\
V1009Cen & Sco-Cen(UCL) & 130 & G8/K0V & 13 & 1.2 & 6.16 & WTT? & 10.18 & 7.95 \\
TYC7310-2431-1 & Sco-Cen(UCL) & 130 & G5 & 16 & 1.2 & 6.16 & WTT & 10.36 & 8.28 \\
TYC8297-1613-1 & Sco-Cen(UCL) & 130 &  & 17 & 1.2 & 6.16 & WTT & 10.22 & 8.51 \\
TYC7822-158-1 & Sco-Cen(UCL) & 130 & K1 & 13 & 1.2 & 6.16 & WTT & 11.11 & 8.51 \\
TYC7848-1659-1 & Sco-Cen(UCL) & 130 & G5 & 15 & 1.2 & 6.16 & WTT & 10.36 & 8.21 \\
HD140421 & Sco-Cen(UCL) & 130 & G1V & 17 & 1.2 & 6.16 & WTT? & 9.46 & 7.87 \\
TYC8317-551-1 & Sco-Cen(UCL) & 130 & G0 & 13 & 1.2 & 6.16 & WTT & 10.29 & 8.27 \\
TYC7333-1260-1 & Sco-Cen(UCL) & 130 & G1/G2V & 18 & 1.2 & 6.16 & WTT & 9.58 & 8.07 \\
PreibZinn9922 &  USco  & 145 & G0 & 18 & 1.1 & 6.03 & … & … & 8.77 \\
TYC9231-1566-1 & Sco-Cen(LCC) & 130 & G3IV & 12 & 1.3 & 5.69 & WTT & 9.23 & 7.18 \\
TYC8263-2453-1 & Sco-Cen(UCL) & 130 & F8/G0V & 14 & 1.3 & 5.69 & WTT & 9.69 & 7.94 \\
TYC7813-224-1 & Sco-Cen(UCL) & 130 &  & 14 & 1.3 & 5.69 & WTT & 10.55 & 8.39 \\
TYC8683-242-1 & Sco-Cen(UCL) & 130 &  & 8 & 1.3 & 5.69 & WTT & 10.80 & 8.30 \\
TYC7828-2913-1 & Sco-Cen(UCL) & 130 &  & 5 & 1.3 & 5.69 & WTT & 11.02 & 8.29 \\
TYC7310-503-1 & Sco-Cen(UCL) & 130 & K3 & 2 & 1.3 & 5.69 & WTT & 10.88 & 7.87 \\
TYC7845-1174-1 & Sco-Cen(UCL) & 130 & K1 & 3 & 1.3 & 5.69 & WTT & 10.61 & 7.93 \\
TYC7349-2191-1 & Sco-Cen(UCL) & 130 &  & 1 & 1.3 & 5.69 & WTT & 11.09 & 8.29 \\
PreibZinn9975 &  USco  & 145 & G9 & 1 & 1.2 & 5.62 & ... & 10.50 & 7.43 \\
HII1514 &  Pleiades  & 130 & G5V & 125 & 1.4 & 5.28 & ... & 10.48 & 8.95 \\
HD140374 & Sco-Cen(UCL) & 130 & G8V & 8 & 1.4 & 5.28 & WTT? & 9.69 & 7.80 \\
CHXR6 &  Cham  & 140 & K2 & 1 & 1.3 & 5.28 & CTT & 11.22 & 7.31 \\
PreibZinn9979 &  USco  & 145 & G5 & 9 & 1.3 & 5.18 & ... &  & 8.69 \\
{\bf High variability targets}  &  &  &  &  &  &  &  &  &  \\
PreibZinn9980 &  USco  & 145 & M1 & 0.3 & 0.3 & 22.87 & … & … & 7.91 \\
PreibZinn9940 &  USco  & 145 & M2 & 0.4 & 0.3 & 22.10 & … & … & 8.61 \\
PreibZinn9970 &  USco  & 145 & M1 & 0.5 & 0.3 & 20.72 & … & … & 8.82 \\
PreibZinn9955 &  USco  & 145 & … & 0.5 & 0.3 & 20.09 & … & … & 8.10 \\
PreibZinn9933 &  USco  & 145 & M1 & 0.8 & 0.4 & 18.95 & … & … & 8.84 \\
\enddata
\tablenotetext{a}{The masses presented here are either taken from the
literature or are estimated using the isochrones of D'Antona and Mazzitelli (1994)}
\tablenotetext{b}{Preibisch \& Zinnecker (1999)}
\end{deluxetable}

\begin{deluxetable}{lccccccccccc}
\tabletypesize{\scriptsize}
\tablecaption{SIM-YSO Sample cont. \label{samtab}}
\tablehead{
\colhead{Name} & \colhead{Cluster} & \colhead{Distance} & \colhead{Spec} & \colhead{Age}   & \colhead{Star Mass$^a$}     & \colhead{Signal}   & \colhead{T-Tauri} & \colhead{V}    & \colhead{2MASS K$_s$} \\
               &                   & \colhead{[pc]}     & \colhead{Type}     & \colhead{[Myr]} & \colhead{[M$_\sun$]}    & \colhead{[$\mu$as]}& \colhead{Class}   & \colhead{[mag]}& \colhead{[mag]} 
}
\startdata
RECX4 &  Eta Cha  & 100 & K7 & 4 & 0.5 & 18.49 & WTT & 12.79 & 8.62 \\
PreibZinn9921 &  USco  & 145 & M1 & 0.8 & 0.4 & 18.42 & … & … & 8.63 \\
PreibZinn993 &  USco  & 145 & M1 & 1 & 0.4 & 17.45 & … & … & 9.08 \\
PreibZinn996 &  USco  & 145 & M0 & 1 & 0.4 & 15.42 & … & … & 8.98 \\
PreibZinn9963 &  USco  & 145 & M0 & 1.8 & 0.5 & 14.74 & … & … & 8.44 \\
DMTau &  Tau Aur  & 140 & K5V:e… & 2 & 0.5 & 14.61 & CTT & 13.78 & 9.52 \\
PreibZinn9959 &  USco  & 145 & M0 & 2 & 0.5 & 14.11 & … & … & 8.91 \\
PreibZinn9962 &  USco  & 145 & M0 & 2 & 0.5 & 14.11 & … & … & 8.92 \\
CHXR29 &  Cham  & 140 & A0pshe & ... & 0.5 & 13.74 & CTT & 8.44 & 5.94 \\
CHXR18N &  Cham  & 140 & K1 & ... & 0.5 & 13.74 & WTT & 12.05 & 7.77 \\
CHXR68A &  Cham  & 140 &  & ... & 0.5 & 13.74 & WTT & 13.37 & 8.87 \\
TCha &  Cham  & 140 & F5 & ... & 0.5 & 13.74 & CTT & 11.86 & 6.95 \\
PreibZinn9911 &  USco  & 145 & K7 & 0.8 & 0.5 & 13.26 & … & … & 8.37 \\
PreibZinn9916 &  USco  & 145 & M0 & 2.9 & 0.5 & 13.26 & … & … & 9.27 \\
DNTau &  Tau Aur  & 140 & K6V:e… & 0.46 & 0.6 & 12.26 & CTT & 12.53 & 8.02 \\
IPTau &  Tau Aur  & 140 & M0:Ve & 2 & 0.6 & 11.84 & CTT & 13.04 & 8.35 \\
PreibZinn9926 &  USco  & 145 & K7 & 1.8 & 0.6 & 11.05 & … & … & 8.92 \\
DGTau &  Tau Aur  & 140 & GV:e… & 2 & 0.7 & 10.57 & CTT & … & 6.99 \\
PreibZinn9961 &  USco  & 145 & K5 & 1.8 & 0.7 & 9.47 & … & … & 8.62 \\
UYAur &  Tau Aur  & 140 & G5V:e… & 2 & 0.7 & 9.28 & CTT & 12.40 & 7.24 \\
BPTau &  Tau Aur  & 140 & K5V:e… & 0.6 & 0.8 & 9.16 & CTT & 11.96 & 7.74 \\
GKTau &  Tau Aur  & 140 &  & 2 & 0.8 & 9.16 & CTT & 12.50 & 7.47 \\
AATau &  Tau Aur  & 140 & M0V:e & 2 & 0.8 & 9.04 & CTT & 12.82 & 8.05 \\
HQTau &  Tau Aur  & 140 &  & 0.691831 & 0.8 & 9.04 & CTT & … & 7.14 \\
IWTau &  Tau Aur  & 140 & K7V & 2 & 0.8 & 9.04 & WTT & 12.51 & 8.28 \\
DRTau &  Tau Aur  & 140 & K4V:e… & 2 & 0.8 & 9.04 & CTT & 13.60 & 6.87 \\
V830Tau &  Tau Aur  & 140 & K7 & 2 & 0.8 & 8.92 & WTT & 12.21 & 8.42 \\
DLTau &  Tau Aur  & 140 & GV:e… & 2 & 0.8 & 8.92 & CTT & 13.55 & 7.96 \\
L1551-51 &  Tau Aur  & 140 & K7 & 2 & ... & 8.59 & … & 12.06 & 8.85 \\
CITau &  Tau Aur  & 140 & GV:e… & 2 & ... & 8.59 & … & 12.99 & 7.79 \\
V836Tau &  Tau Aur  & 140 & K7V & 2.1 & 0.8 & 8.59 & WTT & 13.13 & 8.60 \\
PreibZinn9918 &  USco  & 145 & K3 & 1 & 0.8 & 8.50 & … & … & 8.33 \\
PreibZinn9942 &  USco  & 145 & K5 & 2.9 & 0.8 & 8.50 & … & … & 8.37 \\
PreibZinn9984 &  USco  & 145 & K3 & 1 & 0.8 & 8.50 & … & … & 8.93 \\
TYC8648-446-1 & Sco-Cen(LCC) & 130 & … & 19 & 0.9 & 8.22 & WTT & 11.18 & 8.79 \\
PreibZinn9968 &  USco  & 145 & K2 & 1 & 0.8 & 7.89 & … & 11.65 & 7.69 \\
SR4/V2058Oph &  Ophiuchus  & 160 & K5e & 2 & ... & 7.51 & … & 13.60 & 7.52 \\
DoAr21Oph &  Ophiuchus  & 160 & B2V & 2 & ... & 7.51 & … & 13.82 & 6.23 \\
Haro1-16Oph &  Ophiuchus  & 160 & K3 & 2 & ... & 7.51 & … & 12.59 & 7.61 \\
V1121Oph &  Ophiuchus  & 160 & K5 & 2 & ... & 7.51 & … & 11.25 & 6.96 \\
V966Cen & Sco-Cen(LCC) & 130 & K1:V:+… & ... & 1.0 & 7.40 & WTT? & 9.76 & 8.08 \\
TYC7319-749-1 & Sco-Cen(UCL) & 130 & K0 & 20 & 1.0 & 7.40 & WTT & 10.59 & 8.34 \\
GITau &  Tau Aur  & 140 & K5e & 2 & 0.9 & 7.39 & CTT & 13.50 & 7.89 \\
PreibZinn991 &  USco  & 145 & K3 & 3 & 0.9 & 7.37 & … & … & 9.43 \\
PreibZinn9937 &  USco  & 145 & K4 & 4 & 0.9 & 7.37 & … & … & 8.93 \\
PreibZinn9976 &  USco  & 145 & K1 & 1.2 & 1.0 & 6.98 & … & … & 8.49 \\
V1072Tau/TAP35 &  Tau Aur  & 140 & K1 & 2 & 1.0 & 6.87 & … & 10.30 & 8.30 \\
TYC8640-2515-1 & Sco-Cen(LCC) & 130 & … & 20 & 1.1 & 6.72 & WTT & 10.77 & 8.73 \\
TYC8242-1324-1 & Sco-Cen(LCC) & 130 & G0 & 16 & 1.1 & 6.72 & WTT & 10.38 & 8.14 \\
TYC8238-1462-1 & Sco-Cen(LCC) & 130 & K0 & 21 & 1.1 & 6.72 & WTT & 10.10 & 8.01 \\
TYC8655-149-1 & Sco-Cen(LCC) & 130 & … & 19 & 1.1 & 6.72 & WTT & 10.31 & 8.37 \\
TYC8282-516-1 & Sco-Cen(UCL) & 130 & … & 19 & 1.1 & 6.72 & WTT & 10.68 & 8.54 \\
TYC7833-2559-1 & Sco-Cen(UCL) & 130 & G6/G8III/IV & 21 & 1.1 & 6.72 & WTT & 10.61 & 8.45 \\
TYC8694-1685-1 & Sco-Cen(UCL) & 130 & … & 18 & 1.1 & 6.72 & WTT & 10.21 & 8.01 \\
PreibZinn9944 &  USco  & 145 & K2 & 3.7 & 1.0 & 6.63 & … & … & 8.51 \\
PreibZinn9973 &  USco  & 145 & K0 & 0.8 & 1.0 & 6.57 & … & 11.00 & 7.49 \\
PreibZinn9978 &  USco  & 145 & K0 & 0.3 & 1.0 & 6.50 & … & 10.80 & 7.46 \\
PreibZinn9929 &  USco  & 145 & K3e & 3 & 1.1 & 6.32 & … & 13.40 & 8.52 \\
PreibZinn9954 &  USco  & 145 & M3 & 3 & 1.1 & 6.32 & … & … & 8.86 \\
TYC9246...71-1 & Sco-Cen(LCC) & 130 & G & 7 & 1.2 & 6.16 & CTT & 10.54 & 7.29 \\
TYC9212-2011-1 & Sco-Cen(LCC) & 130 & … & 6 & 1.2 & 6.16 & WTT & 10.49 & 7.79 \\
TYC8644-340-1 & Sco-Cen(LCC) & 130 & … & 13 & 1.2 & 6.16 & WTT & 10.29 & 7.97 \\
TYC8645-1339-1 & Sco-Cen(LCC) & 130 & … & 5 & 1.2 & 6.16 & WTT & 10.82 & 7.73 \\
TYC8249-52-1 & Sco-Cen(LCC) & 130 & K0/K1 & 13 & 1.2 & 6.16 & WTT & 10.48 & 8.13 \\
HD117524 & Sco-Cen(LCC) & 130 & G5/G6V & 15 & 1.2 & 6.16 & WTT? & 9.84 & 7.83 \\
TYC7796-1788-1 & Sco-Cen(UCL) & 130 & K5 & 13 & 1.2 & 6.16 & WTT & 10.17 & 7.88 \\
TYC7811-2909-1 & Sco-Cen(UCL) & 130 & … & 14 & 1.2 & 6.16 & WTT & 10.80 & 8.40 \\
TYC8283-264-1 & Sco-Cen(UCL) & 130 & … & 18 & 1.2 & 6.16 & WTT & 10.09 & 7.90 \\
TYC7824-1291-1 & Sco-Cen(UCL) & 130 & G8IV: & 15 & 1.2 & 6.16 & WTT & 9.80 & 7.81 \\
TYC8294-2230-1 & Sco-Cen(UCL) & 130 & G7 & 17 & 1.2 & 6.16 & WTT & 10.79 & 8.71 \\
TYC7852-51-1 & Sco-Cen(UCL) & 130 & F7V & 18 & 1.2 & 6.16 & WTT & 9.05 & 7.69 \\
PreibZinn9986 &  USco  & 145 & K0 & 1.8 & 1.1 & 6.14 & … & … & 7.76 \\
PreibZinn9983 &  USco  & 145 & K0 & 2 & 1.1 & 6.03 & … & … & 8.51 \\
PreibZinn9971 &  USco  & 145 & K1 & 2.5 & 1.1 & 5.92 & … & 11.65 & 8.09 \\
TYC8982-3213-1 & Sco-Cen(LCC) & 130 & G1/G2V & 13 & 1.3 & 5.69 & WTT & 9.49 & 7.60 \\
HD105070 & Sco-Cen(LCC) & 130 & G1V & 13 & 1.3 & 5.69 & WTT? & 8.89 & 7.31 \\
TYC8234-2856-1 & Sco-Cen(LCC) & 130 & … & 9 & 1.3 & 5.69 & WTT & 10.59 & 8.16 \\
TYC8633-28-1 & Sco-Cen(LCC) & 130 & G2 & 15 & 1.3 & 5.69 & WTT & 9.49 & 7.77 \\
HD108568 & Sco-Cen(LCC) & 130 & G1 & 14 & 1.3 & 5.69 & WTT? & 8.89 & 7.29 \\
HD113466 & Sco-Cen(LCC) & 130 & G5V: & 14 & 1.3 & 5.69 & WTT? & 9.18 & 7.36 \\
TYC7815-2029-1 & Sco-Cen(UCL) & 130 & K0/K1+… & 14 & 1.3 & 5.69 & WTT? & 9.46 & 7.88 \\
\enddata
\tablenotetext{a}{The masses presented here are either taken from the
literature or are estimated using the isochrones of D'Antona and Mazzitelli (1994)}
\tablenotetext{b}{Preibisch \& Zinnecker (1999)}
\end{deluxetable}

\begin{deluxetable}{lccccccccccc}
\tabletypesize{\scriptsize}
\tablecaption{SIM-YSO Sample cont. \label{samtab}}
\tablehead{
\colhead{Name} & \colhead{Cluster} & \colhead{Distance} & \colhead{Spec} & \colhead{Age}   & \colhead{Star Mass$^a$}     & \colhead{Signal}   & \colhead{T-Tauri} & \colhead{V}    & \colhead{2MASS K$_s$} \\
               &                   & \colhead{[pc]}     & \colhead{Type}     & \colhead{[Myr]} & \colhead{[M$_\sun$]}    & \colhead{[$\mu$as]}& \colhead{Class}   & \colhead{[mag]}& \colhead{[mag]} 
}
\startdata
TYC7833-2037-1 & Sco-Cen(UCL) & 130 & K1 & 10 & 1.3 & 5.69 & WTT & 11.23 & 8.73 \\
TYC7840-1280-1 & Sco-Cen(UCL) & 130 & G9 & 9 & 1.3 & 5.69 & WTT & 10.57 & 8.29 \\
PreibZinn9949 &  USco  & 145 & G7 & 8.5 & 1.2 & 5.53 & … & … & 8.46 \\
LkCa19 &  Tau Aur  & 140 & K0V & 2 & 1.3 & 5.49 & WTT & 10.85 & 8.15 \\
PreibZinn9925 &  USco  & 145 & G9 & 4 & 1.2 & 5.44 & … & 10.99 & 8.44 \\
TYC8644-802-1 & Sco-Cen(LCC) & 130 & … & 6 & 1.4 & 5.28 & WTT & 10.21 & 7.66 \\
HD108611 & Sco-Cen(LCC) & 130 & G5V & 10 & 1.4 & 5.28 & WTT? & 9.04 & 7.12 \\
TYC7326...28-1 & Sco-Cen(UCL) & 130 & K1 & 7 & 1.4 & 5.28 & WTT & 10.54 & 8.12 \\
HD138995 & Sco-Cen(UCL) & 130 & G5V & 10 & 1.4 & 5.28 & WTT? & 9.39 & 7.52 \\
TYC7842-250-1 & Sco-Cen(UCL) & 130 & … & 8 & 1.4 & 5.28 & WTT & 10.90 & 8.69 \\
TYC7333-719-1 & Sco-Cen(UCL) & 130 & G8 & 10 & 1.4 & 5.28 & WTT & 10.99 & 8.53 \\
TYC7853-227-1 & Sco-Cen(UCL) & 130 & … & 8 & 1.4 & 5.28 & WTT & 11.05 & 8.65 \\
{\bf Yet to be measured targets }  &  &  &  &  &  &  &  &  &  \\
GJ803 &  Beta Pic  & 9.9 & M1Ve & 20 & 0.4 & 242.81 & … & 8.81 & 4.53 \\
HD155555C &  Beta Pic  & 31.4 & M4.5 & 20 & 0.2 & 153.11 & … & 12.71 & 7.63 \\
HIP23309 &  Beta Pic  & 26.3 & K7V & 20 & 0.5 & 81.25 & … & 10.02 & 6.24 \\
HIP3556 &  Tuc  & 45 & M1.5 & 20 & 0.3 & 71.23 & … & 11.91 & 7.62 \\
GJ3305 &  Beta Pic  & 29.8 & M0.5 & 20 & 0.5 & 64.53 & … & 10.59 & 6.41 \\
CD-64d1208 &  Beta Pic  & 29.2 & M0 & 20 & 0.6 & 54.88 & … & 9.54 & 6.10 \\
GSC8056-0482 &  Hor  & 60 & M3Ve & 30 & 0.3 & 53.42 & … & 12.11 & 7.50 \\
TWA8A &  TW Hya  & 60 & M2 & 10 & 0.3 & 53.42 & … & … & 7.43 \\
TWA10 &  TW Hya  & 60 & M2.5 & 10 & 0.3 & 53.42 & … & … & 8.19 \\
TWA11B &  TW Hya  & 60 & M2.5 & 10 & 0.3 & 53.42 & … & 13.30 & 5.77 \\
HIP107345 &  Tuc  & 45 & M1 & 20 & 0.4 & 53.42 & … & 11.72 & 7.87 \\
AOMen &  Beta Pic  & 38.5 & K3:V: & 20 & 0.6 & 41.63 & … & 9.95 & 6.81 \\
TWA7 &  TW Hya  & 60 & M1 & 10 & 0.4 & 40.06 & … & 11.06 & 6.90 \\
TWA13 &  TW Hya  & 60 & M1Ve & 10 & 0.4 & 40.06 & … & 11.50 & 7.49 \\
HD35850 &  Beta Pic  & 26.8 & F7V & 20 & 1.0 & 35.88 & … & 6.30 & 4.93 \\
HIP1993 &  Tuc  & 45 & K7V & 20 & 0.6 & 35.61 & … & 11.26 & 7.75 \\
GSC8499-0304 &  Hor  & 60 & M0Ve & 30 & 0.5 & 32.05 & … & 12.09 & 8.72 \\
TWA14 &  TW Hya  & 60 & M0 & 10 & 0.5 & 32.05 & … & … & 8.50 \\
TWA18 &  TW Hya  & 60 & M0.5 & 10 & 0.5 & 32.05 & … & … & 8.85 \\
HD3221 &  Tuc  & 45 & K5V & 20 & 0.8 & 26.71 & … & 9.56 & 6.53 \\
GSC8497-0995 &  Hor  & 60 & K6Ve & 30 & 0.6 & 26.71 & … & 10.97 & 7.78 \\
TWA6 &  TW Hya  & 60 & K7 & 10 & 0.6 & 26.71 & … & 12.00 & 8.04 \\
TWA1 &  TW Hya  & 60 & K8Ve & 10 & 0.6 & 26.71 & … & 10.92 & 7.30 \\
TWA19B &  TW Hya  & 60 & K7 & 10 & 0.6 & 26.71 & … & … & 8.28 \\
V343Nor &  Beta Pic  & 39.8 & K0V & 20 & 1.0 & 24.16 & … & 8.14 & 5.85 \\
HD202746 &  Tuc  & 45 & K2Vp… & 20 & 0.9 & 23.74 & … & 8.97 & 6.40 \\
TWA4 &  TW Hya  & 60 & K5 & 10 & 0.7 & 22.89 & … & 8.89 & 5.59 \\
TWA9A &  TW Hya  & 60 & K5 & 10 & 0.7 & 22.89 & … & 11.13 & 7.85 \\
TWA17 &  TW Hya  & 60 & K5 & 10 & 0.7 & 22.89 & … & … & 9.01 \\
HD1466 &  Tuc  & 45 & F8/G0V & 20 & 1.0 & 21.37 & … & 7.46 & 6.15 \\
HD186602 &  Tuc  & 45 & F7/F8V & 20 & 1.0 & 21.37 & … & 7.28 & 6.09 \\
HD207575 &  Tuc  & 45 & F6V & 20 & 1.0 & 21.37 & … & 7.22 & 6.03 \\
PPM366328 &  Tuc  & 45 & K0 & 20 & 1.0 & 21.37 & … & 9.67 & 7.61 \\
GSC8047-0232 &  Hor  & 60 & K3V & 30 & 0.8 & 20.03 & … & 10.87 & 8.41 \\
CD-53d386 &  Hor  & 60 & K3Ve & 30 & 0.8 & 20.03 & … & 11.02 & 8.59 \\
GSC8862-0019 &  Hor  & 60 & K4Ve & 30 & 0.8 & 20.03 & … & 11.67 & 8.91 \\
BetaPic &  Beta Pic  & 19.3 & A5V & 20 & 2.5 & 19.93 & … & 3.85 & 3.53 \\
PZTel &  Beta Pic  & 49.7 & K0Vp & 20 & 1.0 & 19.35 & … & 8.43 & 6.37 \\
HD208233 &  Tuc  & 45 & G8V & 20 & 1.2 & 17.81 & … & 8.90 & 6.75 \\
CCPhe &  Hor  & 60 & K1V & 30 & 0.9 & 17.81 & … & 9.35 & 6.83 \\
CD-65d149 &  Hor  & 60 & K2Ve… & 30 & 0.9 & 17.81 & … & 10.19 & 8.01 \\
HD172555 &  Beta Pic  & 29.2 & A5IV-V & 20 & 2.0 & 16.46 & … & 4.78 & 4.30 \\
HD987 &  Tuc  & 45 & G6V & 20 & 1.3 & 16.44 & … & 8.76 & 6.96 \\
HR9 &  Beta Pic  & 39.1 & F2IV & 20 & 1.5 & 16.39 & … & 6.19 & 5.24 \\
CPD-64120 &  Hor  & 60 & K1Ve & 30 & 1.0 & 16.03 & … & 10.29 & 8.01 \\
HD202917 &  Tuc  & 45 & G5V & 20 & 1.4 & 15.26 & … & 8.65 & 6.91 \\
HD195627 &  Tuc  & 45 & F1III & 20 & 1.5 & 14.25 & … & 4.75 & 4.04 \\
HD195961 &  Tuc  & 45 & Fm… & 20 & 1.5 & 14.25 & … & 4.86 & 3.90 \\
HD164249 &  Beta Pic  & 46.9 & F5V & 20 & 1.5 & 13.67 & … & 7.01 & 5.91 \\
HD207129 &  Tuc  & 45 & G0V & 20 & 1.6 & 13.35 & … & 5.57 & 4.24 \\
IQTau &  Tau Aur  & 140 & M2 & 2 & 0.5 & 13.21 & CTT & 14.50 & 7.78 \\
HD181327 &  Beta Pic  & 50.6 & F5/F6V & 20 & 1.5 & 12.67 & … & 7.04 & 5.91 \\
HD178085 &  Tuc  & 45 & G0V & 20 & 1.7 & 12.57 & … & 8.31 & 6.88 \\
RXJ012320.9-572853 &  HOR  & 60 & G6V & 30 & 1.3 & 12.33 & … & 8.53 & 6.85 \\
DQTau &  Tau Aur  & 140 & M0V:e… & 2 & 0.6 & 12.05 & CTT & 13.66 & 7.98 \\
RXJ020718.6-531155 &  HOR  & 60 & G5V & 30 & 1.4 & 11.45 & … & 8.64 & 6.89 \\
TWA19A &  TW Hya  & 60 & G3/G5Vp & 10 & 1.4 & 11.45 & … & 9.07 & 7.51 \\
RXJ020436.7-545320 &  HOR  & 60 & F2V & 30 & 1.5 & 10.68 & … & 6.45 & 5.45 \\
HD200798 &  Tuc  & 45 & A5/A6IV/V & 20 & 2.0 & 10.68 & … & 6.69 & 6.07 \\
GMTau &  Tau Aur  & 140 & M6.5 & 2 & ... & 8.59 & … &  & 10.63 \\
I045251+3016 &  Tau Aur  & 140 & K5 & 2 & ... & 8.59 & … & 11.60 & 8.13 \\
Haro6-37 &  Tau Aur  & 140 & K6 & 2 & 0.8 & 8.48 & CTT & 13.42 & 7.31 \\
VXR03 &  IC2391  & 155 & … & 53 & ... & 7.75 & … & 10.95 & 14.86 \\
L36 &  IC2391  & 155 & F6V & 53 & ... & 7.75 & … & 9.83 & 8.63 \\
VXR31 &  IC2391  & 155 & … & 53 & ... & 7.75 & … & 11.22 & 9.69 \\
H21 &  IC2391  & 155 & … & 53 & ... & 7.75 & … & 11.69 & 9.54 \\
SHJM3 &  IC2391  & 155 & K3e & 53 & 0.8 & 7.75 & … & 12.63 & 9.69 \\
L33 &  IC2391  & 155 & F5V & 53 & ... & 7.75 & … & 9.59 & 8.36 \\
H35 &  IC2391  & 155 & F9 & 53 & ... & 7.75 & … & 10.34 & 8.99 \\
CHXR3 &  Cham  & 140 & K3 & 0.5 & 0.9 & 7.63 & WTT & 12.26 & 7.36 \\
CHXR10 &  Cham  & 140 & M0 & 2 & 0.9 & 7.63 & CTT & 11.69 & 8.20 \\
\enddata
\tablenotetext{a}{The masses presented here are either taken from the
literature or are estimated using the isochrones of D'Antona and Mazzitelli (1994)}
\tablenotetext{b}{Preibisch \& Zinnecker (1999)}
\end{deluxetable}

\begin{deluxetable}{lccccccccccc}
\tabletypesize{\scriptsize}
\tablecaption{SIM-YSO Sample cont. \label{samtab}}
\tablehead{
\colhead{Name} & \colhead{Cluster} & \colhead{Distance} & \colhead{Spec} & \colhead{Age}   & \colhead{Star Mass$^a$}     & \colhead{Signal}   & \colhead{T-Tauri} & \colhead{V}    & \colhead{2MASS K$_s$} \\
               &                   & \colhead{[pc]}     & \colhead{Type}     & \colhead{[Myr]} & \colhead{[M$_\sun$]}    & \colhead{[$\mu$as]}& \colhead{Class}   & \colhead{[mag]}& \colhead{[mag]} 
}
\startdata
ROX3/V2245Oph &  Ophiuchus  & 160 & M1 & 2 & ... & 7.51 & … & 13.12 & 8.78 \\
HR6070 &  Beta Pic  & 43 & A0V & 20 & 3.0 & 7.45 & … & 4.80 & 4.74 \\
PreibZinn9985 &  USco  & 145 & K2 & 2 & 0.9 & 7.13 & … & … & 8.18 \\
HR136 &  Tuc  & 45 & A0V & 20 & 3.0 & 7.12 & … & 5.07 & 4.99 \\
HR9062 &  Tuc  & 45 & A1V & 20 & 3.0 & 7.12 & … & 5.00 & 4.82 \\
SHJM6 &  IC2391  & 155 & K0 & 53 & 1.0 & 6.20 & … & 11.86 & 9.79 \\
VXR62 &  IC2391  & 155 & … & 53 & 1.0 & 6.20 & … & 11.73 & 15.03 \\
VXR67 &  IC2391  & 155 & … & 53 & 1.0 & 6.20 & … & 11.71 & 13.61 \\
VXR16 &  IC2391  & 155 & … & 53 & 1.1 & 5.64 & … & 11.84 & 14.58 \\
VXR72 &  IC2391  & 155 & G9 & 53 & 1.1 & 5.64 & … & 11.46 & 9.59 \\
HR126 &  Tuc  & 45 & B9V & 20 & 4.0 & 5.34 & … & 4.36 & 4.48 \\
TWA11A &  TW Hya  & 60 & A0V & 10 & 3.0 & 5.34 & … & 5.78 & 5.77 \\
\enddata
\tablenotetext{a}{The masses presented here are either taken from the
literature or are estimated using the isochrones of D'Antona and Mazzitelli (1994)}
\tablenotetext{b}{Preibisch \& Zinnecker (1999)}
\end{deluxetable}

\begin{deluxetable}{lccc|lccc}
\tabletypesize{\scriptsize}
\tablecaption{Stellar Variabilities \label{phottab}}
\tablehead{ \colhead{Target} & \colhead{Peak to Peak} & \colhead{$\sigma_{stdev}$} & \colhead{Program} & \colhead{Target} & \colhead{Peak to Peak} & \colhead{$\sigma_{stdev}$} & \colhead{Program}
}
\startdata
 HII1215    &   0.024   &   0.009   &   Maidanak   &   PreibZinn99 68    &   0.140   &   0.045   &   SMARTS   \\
51 Eri    &   0.019   &   0.010   &   Maidanak   &   TYC7326-928-1    &   0.140   &   0.045   &   SMARTS   \\
 GJ3305    &   0.020   &   0.010   &   Maidanak   &    5251+3060    &   0.150   &   0.045   &   Maidanak   \\
HII996    &   0.030   &   0.010   &   Maidanak   &    TYC8238-1462-1    &   0.150   &   0.045   &   SMARTS   \\
 HII489    &   0.029   &   0.012   &   Maidanak   &   CHXR 65    &   0.170   &   0.045   &   SMARTS   \\
 HII1309    &   0.037   &   0.013   &   Maidanak   &   DM Tau    &   0.136   &   0.046   &   Maidanak   \\
CHXR 11    &   0.040   &   0.013   &   SMARTS   &    PreibZinn99 59    &   0.140   &   0.046   &   SMARTS   \\
 HD141569    &   0.056   &   0.013   &   Maidanak   &    TYC7319-749-1    &   0.180   &   0.046   &   SMARTS   \\
HII1207    &   0.034   &   0.014   &   Maidanak   &   PreibZinn99 61    &   0.140   &   0.047   &   SMARTS   \\
 HII1797    &   0.044   &   0.014   &   Maidanak   &    PreibZinn99 1    &   0.140   &   0.047   &   SMARTS   \\
HII1856    &   0.051   &   0.014   &   Maidanak   &    PreibZinn99 29    &   0.170   &   0.047   &   SMARTS   \\
 HII1095    &   0.050   &   0.015   &   Maidanak   &   PreibZinn99 37    &   0.140   &   0.048   &   SMARTS   \\
 HII1613    &   0.054   &   0.015   &   Maidanak   &   TYC8655-149-1    &   0.140   &   0.048   &   SMARTS   \\
 HD140374    &   0.060   &   0.015   &   SMARTS   &    TYC7853-227-1    &   0.140   &   0.048   &   SMARTS   \\
 HII2366    &   0.041   &   0.016   &   Maidanak   &    PreibZinn99 32    &   0.130   &   0.049   &   SMARTS   \\
 PreibZinn99 69    &   0.050   &   0.016   &   SMARTS   &   TYC8283-264-1    &   0.150   &   0.049   &   SMARTS   \\
HII1794    &   0.055   &   0.016   &   Maidanak   &    HD149735    &   0.130   &   0.050   &   SMARTS   \\
TYC9244-814-1    &   0.050   &   0.017   &   SMARTS   &    TYC7840-1280-1    &   0.170   &   0.051   &   SMARTS   \\
TYC8259-689-1    &   0.060   &   0.017   &   SMARTS   &   TYC7842-250-1    &   0.130   &   0.052   &   SMARTS   \\
HII430    &   0.048   &   0.018   &   Maidanak   &    TYC8644-340-1    &   0.130   &   0.052   &   SMARTS   \\
TYC8270-2015-1    &   0.050   &   0.018   &   SMARTS   &    TYC8694-1685-1    &   0.150   &   0.052   &   SMARTS   \\
HII1275    &   0.056   &   0.018   &   Maidanak   &    PreibZinn99 76    &   0.160   &   0.052   &   SMARTS   \\
 TYC7871-1282-1    &   0.070   &   0.018   &   SMARTS   &    TYC7833-2037-1    &   0.800   &   0.052   &   SMARTS   \\
HII1514    &   0.059   &   0.019   &   Maidanak   &   TYC7852-51-1    &   0.140   &   0.053   &   SMARTS   \\
HD149551    &   0.060   &   0.019   &   SMARTS   &   PreibZinn99 26    &   0.170   &   0.054   &   SMARTS   \\
 PreibZinn99 74    &   0.060   &   0.019   &   SMARTS   &   PreibZinn99 16    &   0.170   &   0.055   &   SMARTS   \\
TYC8646-166-1    &   0.070   &   0.019   &   SMARTS   &   PreibZinn99 84    &   0.170   &   0.055   &   SMARTS   \\
 PreibZinn9 10    &   0.070   &   0.019   &   SMARTS   &    PreibZinn99 86    &   0.190   &   0.056   &   SMARTS   \\
TYC7848-1659-1    &   0.060   &   0.020   &   SMARTS   &    TYC7824-1291-1    &   0.190   &   0.057   &   SMARTS   \\
 CHXR 37    &   0.070   &   0.020   &   SMARTS   &   PreibZinn99 23    &   0.150   &   0.058   &   SMARTS   \\
 CHXR 8    &   0.070   &   0.020   &   SMARTS   &   TYC8249-52-1    &   0.150   &   0.058   &   SMARTS   \\
 PreibZinn99 19    &   0.070   &   0.021   &   SMARTS   &    TYC8645-1339-1    &   0.170   &   0.058   &   SMARTS   \\
 TYC7353-2640-1    &   0.080   &   0.021   &   SMARTS   &   HD108611    &   0.180   &   0.058   &   SMARTS   \\
RECX5    &   0.090   &   0.021   &   SMARTS   &   TYC7811-2909-1    &   0.150   &   0.062   &   SMARTS   \\
 TYC9245-617-1    &   0.060   &   0.022   &   SMARTS   &   TYC9212-2011-1    &   0.160   &   0.062   &   SMARTS   \\
PreibZinn99 75    &   0.070   &   0.022   &   SMARTS   &   ROX3    &   0.180   &   0.062   &   SMARTS   \\
TYC7828-2913-1    &   0.070   &   0.022   &   SMARTS   &   V1121 Oph    &   0.270   &   0.063   &   SMARTS   \\
 PreibZinn99 64    &   0.070   &   0.022   &   SMARTS   &   TYC7815-2029-1    &   0.220   &   0.064   &   SMARTS   \\
TYC8633-508-1    &   0.080   &   0.022   &   SMARTS   &   TYC7858-830-1    &   0.170   &   0.065   &   SMARTS   \\
 TYC8636-2515-1    &   0.090   &   0.022   &   SMARTS   &   PreibZinn99 6    &   0.190   &   0.065   &   SMARTS   \\
 PreibZinn99 81    &   0.080   &   0.023   &   SMARTS   &   PreibZinn99 63    &   0.200   &   0.066   &   SMARTS   \\
HII1032    &   0.068   &   0.024   &   Maidanak   &   TAP35    &   0.308   &   0.066   &   Maidanak   \\
 TYC7310-2431-1    &   0.080   &   0.024   &   SMARTS   &    TYC7796-1788-1    &   0.170   &   0.067   &   SMARTS   \\
 TYC8283-2795-1    &   0.080   &   0.024   &   SMARTS   &   PreibZinn99 40    &   0.180   &   0.067   &   SMARTS   \\
 TYC8667-283-1    &   0.090   &   0.024   &   SMARTS   &    RECX4    &   0.200   &   0.067   &   SMARTS   \\
TYC7305-380-1    &   0.090   &   0.025   &   SMARTS   &   PreibZinn99 77    &   0.160   &   0.068   &   SMARTS   \\
TYC7349-2191-1    &   0.090   &   0.025   &   SMARTS   &   PreibZinn99 3    &   0.200   &   0.068   &   SMARTS   \\
TYC7310-503-1    &   0.080   &   0.026   &   SMARTS   &    CHXR 18N    &   0.210   &   0.068   &   SMARTS   \\
PreibZinn99 66    &   0.080   &   0.027   &   SMARTS   &    SR4    &   0.200   &   0.069   &   SMARTS   \\
PreibZinn99 82    &   0.080   &   0.027   &   SMARTS   &    TYC7333-719-1    &   0.240   &   0.069   &   SMARTS   \\
TYC8297-1613-1    &   0.100   &   0.027   &   SMARTS   &    ROX43A    &   0.180   &   0.072   &   SMARTS   \\
 TYC9231-1566-1    &   0.100   &   0.027   &   SMARTS   &    Haro6-37    &   0.230   &   0.075   &   SMARTS   \\
PreibZinn99 2    &   0.090   &   0.028   &   SMARTS   &    CHXR 66    &   0.220   &   0.076   &   SMARTS   \\
 TYC8258-1878-1    &   0.090   &   0.028   &   SMARTS   &   PreibZinn99 18    &   0.240   &   0.077   &   SMARTS   \\
 TYC8317-551-1    &   0.100   &   0.028   &   SMARTS   &    PreibZinn99 55    &   0.230   &   0.078   &   SMARTS   \\
TYC8652-1791-1    &   0.100   &   0.029   &   SMARTS   &    V830 Tau    &   0.346   &   0.078   &   Maidanak   \\
 TYC7845-1174-1    &   0.110   &   0.029   &   SMARTS   &   PreibZinn99 70    &   0.230   &   0.079   &   SMARTS   \\
CHXR 40    &   0.100   &   0.030   &   SMARTS   &   TYC8294-2230-1    &   0.230   &   0.079   &   SMARTS   \\
 PreibZinn99 13    &   0.100   &   0.030   &   SMARTS   &   L1551-51    &   0.258   &   0.079   &   Maidanak   \\
 PreibZinn99 45    &   0.100   &   0.030   &   SMARTS   &    V966 Cen    &   0.130   &   0.080   &   SMARTS   \\
TYC7783-1908-1    &   0.110   &   0.030   &   SMARTS   &   PreibZinn99 11    &   0.230   &   0.081   &   SMARTS   \\
 PreibZinn99 17    &   0.110   &   0.030   &   SMARTS   &   HD105070    &   0.230   &   0.083   &   SMARTS   \\
 TYC8983-98-1    &   0.110   &   0.030   &   SMARTS   &   CHXR 29    &   0.260   &   0.083   &   SMARTS   \\
PreibZinn99 58    &   0.090   &   0.031   &   SMARTS   &    TYC7813-224-1    &   0.080   &   0.084   &   SMARTS   \\
 TYC8295-1530-1    &   0.100   &   0.031   &   SMARTS   &   TYC8982-3213-1    &   0.250   &   0.085   &   SMARTS   \\
TYC7822-158-1    &   0.110   &   0.031   &   SMARTS   &    IWTau    &   0.238   &   0.086   &   Maidanak   \\
 TYC7333-1260-1    &   0.110   &   0.031   &   SMARTS   &    V1056 Sco    &   0.230   &   0.087   &   SMARTS   \\
V1009 Cen    &   0.090   &   0.032   &   SMARTS   &   TYC9246-971-1    &   0.240   &   0.088   &   SMARTS   \\
 PreibZinn99 39    &   0.090   &   0.034   &   SMARTS   &    PreibZinn99 25    &   0.250   &   0.088   &   SMARTS   \\
HD140421    &   0.110   &   0.034   &   SMARTS   &   PreibZinn99 33    &   0.250   &   0.091   &   SMARTS   \\
 TYC8263-2453-1    &   0.110   &   0.034   &   SMARTS   &   PreibZinn99 54    &   0.240   &   0.092   &   SMARTS   \\
PreibZinn99 14    &   0.120   &   0.034   &   SMARTS   &   PreibZinn99 49    &   0.210   &   0.093   &   SMARTS   \\
LkCa19    &   0.163   &   0.034   &   Maidanak   &   TYC8242-1324-1    &   0.320   &   0.094   &   SMARTS   \\
 TYC8654-1115-1    &   0.120   &   0.035   &   SMARTS   &   PreibZinn99 80    &   0.260   &   0.097   &   SMARTS   \\
TYC8640-2515-1    &   0.130   &   0.035   &   SMARTS   &    PreibZinn99 42    &   0.290   &   0.097   &   SMARTS   \\
PreibZinn99-79    &   0.100   &   0.036   &   SMARTS   &    PreibZinn99 21    &   0.300   &   0.107   &   SMARTS   \\
TYC8683-242-1    &   0.110   &   0.036   &   SMARTS   &    HQTau    &   0.233   &   0.109   &   Maidanak   \\
PreibZinn99 28    &   0.110   &   0.037   &   SMARTS   &   CI Tau    &   0.298   &   0.114   &   Maidanak   \\
 PreibZinn99 60    &   0.110   &   0.037   &   SMARTS   &   IPTau    &   0.362   &   0.125   &   Maidanak   \\
 PreibZinn99 67    &   0.120   &   0.037   &   SMARTS   &    PreibZinn99 15    &   0.430   &   0.126   &   SMARTS   \\
HII1124    &   0.100   &   0.038   &   Maidanak   &    TYC7817-622-1    &   0.380   &   0.127   &   SMARTS   \\
\enddata
\end{deluxetable}

\begin{deluxetable}{lccc|lccc}
\tabletypesize{\scriptsize}
\tablecaption{Stellar Variabilities \label{phottab}}
\tablehead{ \colhead{Target} & \colhead{Peak to Peak} & \colhead{$\sigma_{stdev}$} & \colhead{Program} & \colhead{Target} & \colhead{Peak to Peak} & \colhead{$\sigma_{stdev}$} & \colhead{Program}
}
\startdata
PreibZinn99 27    &   0.140   &   0.038   &   SMARTS   &    BP Tau    &   0.457   &   0.129   &   Maidanak   \\
 PreibZinn99 62    &   0.140   &   0.038   &   SMARTS   &    UYAur    &   0.350   &   0.130   &   Maidanak   \\
 RECX6    &   0.150   &   0.038   &   SMARTS   &   SR10    &   0.370   &   0.131   &   SMARTS   \\
 DN Tau    &   0.165   &   0.038   &   Maidanak   &   DoAr 21    &   0.310   &   0.134   &   SMARTS   \\
 PreibZinn99 36    &   0.110   &   0.039   &   SMARTS   &    DG Tau    &   0.606   &   0.138   &   Maidanak   \\
 CHXR 6    &   0.120   &   0.039   &   SMARTS   &    PreibZinn99 78    &   0.350   &   0.139   &   SMARTS   \\
 HD120411    &   0.120   &   0.039   &   SMARTS   &   V836 Tau    &   0.417   &   0.142   &    Maidanak    \\
 RECX10    &   0.120   &   0.039   &   SMARTS   &    HD113466    &   0.390   &   0.149   &   SMARTS   \\
 TYC8248-539-1    &   0.120   &   0.039   &   SMARTS   &   GK Tau    &   0.692   &   0.183   &   Maidanak   \\
PreibZinn9 22    &   0.110   &   0.040   &   SMARTS   &   PreibZinn99 73    &   0.550   &   0.188   &   SMARTS   \\
TYC8319-1687-1    &   0.120   &   0.040   &   SMARTS   &    DL Tau    &   0.723   &   0.212   &   Maidanak   \\
 PreibZinn99 71    &   0.140   &   0.040   &   SMARTS   &   DH Tau    &   0.537   &   0.214   &   Maidanak   \\
HD138995    &   0.150   &   0.040   &   SMARTS   &   GI Tau    &   0.597   &   0.216   &   Maidanak   \\
 L1551-55    &   0.126   &   0.041   &   Maidanak   &    TYC8633-28-1    &   0.180   &   0.226   &   SMARTS   \\
TYC7833-2559-1    &   0.140   &   0.041   &   SMARTS   &    HD108568    &   0.550   &   0.233   &   SMARTS   \\
 PreibZinn99 83    &   0.130   &   0.042   &   SMARTS   &   PreibZinn99 44    &   0.670   &   0.262   &   SMARTS   \\
TWA19    &   0.140   &   0.042   &   SMARTS   &    DR Tau    &   1.092   &   0.296   &   Maidanak   \\
HD117524    &   0.140   &   0.043   &   SMARTS   &   AA Tau    &   1.499   &   0.454   &   Maidanak   \\
TYC8644-802-1    &   0.140   &   0.043   &   SMARTS   &   TYC8234-2856-1    &   0.170   &   0.495   &   SMARTS   \\
CHXR 68A    &   0.150   &   0.043   &   SMARTS   &    TYC8282-516-1    &   0.130   &   0.689   &   SMARTS   \\
 TYC7851-1-1    &   0.120   &   0.044   &   SMARTS   &   RECX2    &   2.150   &   0.808   &   SMARTS   \\
 TYC8648-446-1    &   0.150   &   0.044   &   SMARTS   &    TCha    &   2.430   &   0.850   &   SMARTS   \\
 PreibZinn99 50    &   0.120   &   0.045   &   SMARTS   &    HII1136    &   1.841   &   1.127   &   Maidanak   \\
\enddata
\end{deluxetable}

\clearpage

\begin{figure}[ht]
\plotone{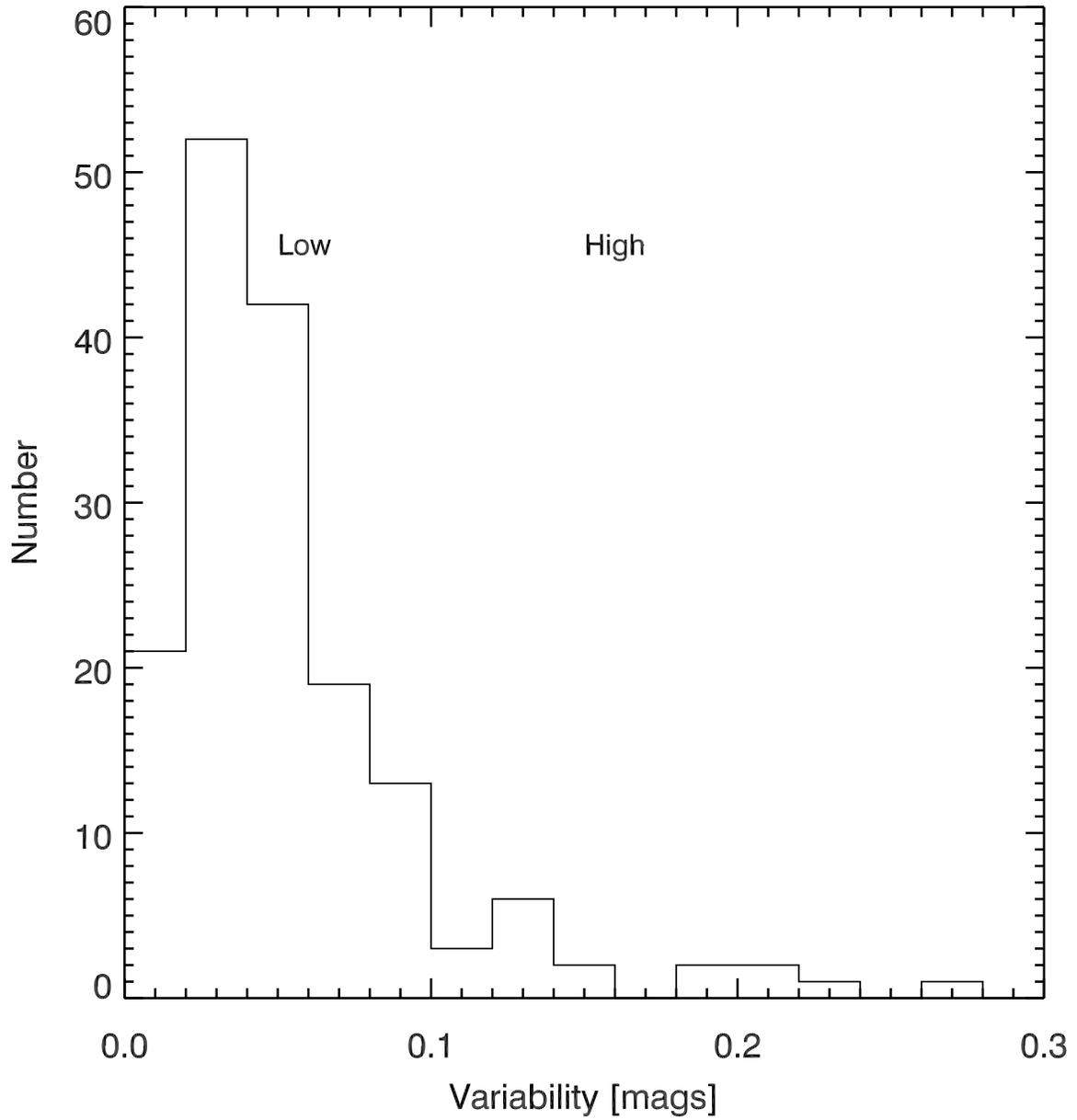} 
\figcaption{Histograms of the standard deviations of the flux variations observed in both the
Northern and Southern photometry surveys. A deviation of $>$ 0.05 magnitudes
is considered too high for the SIM-YSO targets. \label{phothist}}
\end{figure}


\begin{references}

\reference{} Akeson, R.~L., et al.\ 2005, \apj, 635, 1173 

\reference{} Baraffe, I., Chabrier, G., Barman, T.~S., Allard, F., \& Hauschildt, P.~H.\ 2003, \aap, 402, 701 

\reference{} Beichman, C. A. 2001 in {\it Young Stars Near Earth: Progress and Prospects},  
ASP Conference Series Vol. 244. Edited by Ray Jayawardhana and Thomas Greene. 
San Francisco: Astronomical Society of the Pacific,  p.376

\reference{} Boss, A.~P.\ 2001, \apj, 563, 367 

\reference{} Bouvier, J., Rigaut, F., \& Nadeau, D.\ 1997, \aap, 323, 139 

\reference{} Bouvier, J., Covino, 
E., Kovo, O., Martin, E.~L., Matthews, J.~M., Terranegra, L., \& Beck, 
S.~C.\ 1995, \aap, 299, 89 

\reference{} Bouvier, J., \& Bertout, C.\ 1989, \aap, 211, 99 

\reference{} Brandner, W., et al.\ 2000, \aj, 120, 950 

\reference{} Burrows, A., et al.\ 1997, \apj, 491, 856 

\reference{} Butler, R.~P., et al.\  2006, \apj, 646, 505 

\reference{} Catanzarite, J., Shao, M., Tanner, A., Unwin, S., \& Yu, J.\ 2006, \pasp, 118, 1322 

\reference{} Chauvin, G., et al.\ 2005, \aap, 438, L29 

\reference{} Chauvin, G., Lagrange, A.-M., Dumas, C., Zuckerman, B., Mouillet, D., Song, I., Beuzit, J.-L., \& Lowrance, P.\ 2004, \aap, 425, L29 

\reference{} Colavita, M., et al.\ 2003, \apjl, 592, L83 

\reference{} Cutri, R., et al., 2006, AJ, 31, 1163

\reference{} Eiroa, C., et al.\ 2002, \aap, 384, 1038 

\reference{} Frink, S., R{\"o}ser, S., Neuh{\"a}user, R., \& Sterzik, M.~F.\ 1997, \aap, 325, 613 

\reference{} Ghez, A.~M., Neugebauer, G., \& Matthews, K.\ 1993, \aj, 106, 2005 

\reference{} Grankin, K.~N., Melnikov, S.~Y., Bouvier, J., Herbst, W., \& Shevchenko, V.~S.\ 2007, \aap, 
461, 183 

\reference{} Hayward, T.~L., Brandl, B., Pirger, B., Blacken, C., Gull, G.~E., Schoenwald, J., \& Houck, J.~R.\ 
2001, \pasp, 113, 105 

\reference{} Herbst, W., Herbst, D.~K., Grossman, E.~J., \& Weinstein, D.\ 1994, \aj, 108, 1906 

\reference{} Huerta, M., et al. 2007, \apj, in prep

\reference{}  Huerta, M., Prato, L., 
Hartigan, P., Johns-Krull, C.~M., \& Jaffe, D.\ 2005, Bulletin of the 
American Astronomical Society, 37, 1267 

\reference{} Kenyon, S.~J., Dobrzycka, D., \& Hartmann, L.\ 1994, \aj, 108, 1872 

\reference{} Konopacky, Q.~M., Ghez, A.~M., Rice, E.~L., \& Duchene, G.\ 2007, ArXiv Astrophysics 
e-prints, arXiv:astro-ph/0703567 

\reference{} Krist, J.~E., Stapelfeldt, K.~R., M{\'e}nard, F., Padgett, D.~L., \& Burrows, C.~J.\ 
2000, \apj, 538, 793 

\reference{} Ida, S., \& Lin, D.~N.~C.\ 2004, \apj, 616, 567 

\reference{} Law, N.~M., Hodgkin, S.~T., \& Mackay, C.~D.\ 2006, \mnras, 368, 1917 

\reference{} Lin, D.\ 2001, ASP Conf.~Ser.~245: Astrophysical Ages and Times Scales, 245, 90 

\reference{} Lowrance, P.~J., et al.\ 2005, \aj, 130, 1845  

\reference{} Marcy, G.~W., \& Butler, R.~P.\ 2000, \pasp, 112, 137 

\reference{} Mathieu, R.~D., Stassun, K., Basri, G., Jensen, E.~L.~N., Johns-Krull, C.~M., Valenti, J.~A., \& Hartmann, L.~W.\ 1997, \aj, 113, 1841 

\reference{} Mekkaden, M.~V.\ 1998, \aap, 340, 135 

\reference{} Metchev, S., 2005, PhD Thesis, California Institute of Technology

\reference{} Metchev, S.~A., \& Hillenbrand, L.~A.\ 2004, \apj, 617, 1330

\reference{} Neuh{\"a}user, 
R., Guenther, E.~W., Wuchterl, G., Mugrauer, M., Bedalov, A., \& 
Hauschildt, P.~H.\ 2005, \aap, 435, L13 

\reference{} Pan, S., Shao, M., Kulkarni, S., 2004, Nature, 427, 326

\reference{} Preibisch, T., \& Zinnecker, H.\ 1999, \aj, 117, 2381 

\reference{} Pollack, J.~B., 
Hubickyj, O., Bodenheimer, P., Lissauer, J.~J., Podolak, M., \& Greenzweig, 
Y.\ 1996, Icarus, 124, 62 

\reference{} Schneider, J., 2007, http://exoplanet.eu/

\reference{} Schuessler, M., 
Caligari, P., Ferriz-Mas, A., Solanki, S.~K., \& Stix, M.\ 1996, \aap, 314, 
503 

\reference{} Song, I., Zuckerman, B., \& Bessell, M.~S.\ 2003, \apj, 599, 342 

\reference{} Sozzetti, A., 
Casertano, S., Brown, R.~A., \& Lattanzi, M.~G.\ 2003, \pasp, 115, 1072  

\reference{} Stauffer, J.~R., Schild, R., Barrado y Navascues, D., Backman, D.~E., Angelova, A.~M., 
Kirkpatrick, J.~D., Hambly, N., \& Vanzi, L.\ 1998, \apj, 504, 805 

\reference{} Steffen, A.~T., et al.\ 2001, \aj, 122, 997 

\reference{} Strassmeier, K.~G., \& Rice, J.~B.\ 1998, \aap, 339, 497 

\reference{} Unwin, S.~C.\ 2005, Astrometry in the Age of the Next Generation of Large Telescopes, 338, 37 

\reference{} White, R.~J., \& Ghez, A.~M.\ 2001, \apj, 556, 265 

\reference{} Wuchterl, G., \& Tscharnuter, W.~M.\ 2003, \aap, 398, 1081 

\end{references}
\end{document}